\documentclass[
twocolumn,
amsmath,
amssymb,
floatfix,
nofootinbib,
aps,
prb,
longbibliography
]{revtex4-2}

\usepackage[english]{babel}
\usepackage{hyperref}
\usepackage[T1]{fontenc}
\usepackage[utf8]{inputenc}
\usepackage{textcomp} 
\usepackage{lmodern} 
\usepackage{amsmath}
\usepackage{xcolor}

\usepackage{txfonts}
\usepackage{amsfonts}
\usepackage{graphicx}
\graphicspath{{./figures/final/}}

\hypersetup{
    breaklinks = true,
    colorlinks = true,
    pdftitle = {}, 
    pdfauthor = {},
    pdfkeywords = {},
    linkcolor = blue,
    citecolor = blue,
    filecolor = black,
    urlcolor = blue
}

\newcommand{\loz}{L1$\mathrm{_0}$} 
\newcommand{\fecoc}{(Fe$_{1-x}$Co$_x$)$_{16}$C} 
\newcommand{\feco}{Fe$_{1-x}$Co$_x$}

\newcommand{\fecon}{(Fe$_{1-x}$Co$_x$)$_{16}$N$_2$}
\newcommand{\muevat}{$\muup$eV\,atom$^{-1}$}

\newcommand{\mub}{$\muup_{\rm B}$}
\newcommand{\mubat}{$\muup_{\rm B}$\,atom$^{-1}$}
\newcommand{\mjmqb}{MJ\,m$^{-3}$}
\newcommand{\ku}{$K\rm _u$}

\newcommand{\etal}{\textit{et~al.}}

\begin{document}
\begin{sloppypar} 

\title{Structural and magnetic properties of Fe-Co-C alloys with tetragonal deformation: a first-principle study}

\author{Wojciech Marciniak}%
	\email[email: ]{wojciech.robe.marciniak@doctorate.put.poznan.pl}%
	\affiliation{Institute of Molecular Physics, Polish Academy of Sciences,  M. Smoluchowskiego 17, 60-179 Pozna\'n, Poland}%
	\affiliation{Institute of Physics, Poznan University of Technology, Piotrowo 3, 60-965 Pozna\'n, Poland}%
\author{Miros\l{}aw Werwi\'nski}%
        \email[email: ]{werwinski@ifmpan.poznan.pl}%
	\affiliation{Institute of Molecular Physics, Polish Academy of Sciences,  M. Smoluchowskiego 17, 60-179 Pozna\'n, Poland}

\begin{abstract}

%
Fe\nobreakdash-Co alloys with induced tetragonal strain are promising materials for rare-earth-free permanent magnets.
%
%
However, as ultrathin-film studies have shown, tetragonal Fe\nobreakdash-Co structures tend to a rapid relaxation toward a cubic structure as the thickness of the deposited film increases.
One of the main methods of inducing the stable strain in the bulk material is interstitial doping with small atoms, like B, C, or N.
%
%
In this work, we present a full configuration space analysis in the density functional theory approach for \fecoc{} supercells with a single C impurity in one of the octahedral interstitial positions and for the full range of Co concentrations $x$. 
%
%
We discuss all assumptions and considerations leading to calculated lattice parameters, mixing enthalpies, magnetic moments, and averaged magnetocrystalline anisotropy energies (MAE).
We present a comprehensive qualitative analysis of the structural and magnetic properties dependence on short- and long-range ordering parameters. 
%
%
We analyzed all unique Fe/Co atoms occupancies at all stoichiometric concentrations possible in 2~$\times$~2~$\times$~2 supercell based on 2-atom tetragonal representation.
We rely on the thermodynamic averaging method and large sample count to obtain accurate MAE values.
%
%
We place the utilized method in the context of several chemical disorder approximation methods, including effective medium methods (virtual crystal approximation and coherent potential approximation) and special quasirandom structures method applied to Fe\nobreakdash-Co\nobreakdash-based alloys.
We observe a structural phase transition from the body-centered tetragonal structure above 70\% Co concentration and confirm the structural stability of Fe\nobreakdash-Co\nobreakdash-C alloys in the tetragonal range.
We show the presence of a broad MAE maximum around about 50\% Co concentration and notably high MAE values for Co content $x$ as low as 25\%. 
In addition, we show the presence of a positive correlation between MAE and mixing enthalpy.

\end{abstract}


\maketitle

\section{Introduction}


Permanent magnets are an indispensable part of modern technology.
Among their main characteristic parameters are the energy product $(BH)_{\rm max}$ and coercive field $H_{\rm C}$. 
$(BH)_{\rm max}$ determines the efficiency of a permanent magnet and mainly depends on the saturation magnetization $M_{\rm S}$ and coercive field.
Most of the current high-end magnets, with outstanding performance, contain rare-earth elements, such as samarium in SmCo$_5$ and neodymium in Nd$_2$Fe$_{14}$B. 
However, rare-earth-based magnets have limitations, such as the relatively low Curie temperature of neodymium magnets, which is insufficient for many applications.
Moreover, concerns have risen recently about the rare-earth market fragility, which manifested in the so-called rare-earth crisis in 2011~\cite{coey_permanent_2012,gutfleisch_magnetic_2011}. 
Hence, intense research for rare-earth-free permanent magnets has been conducted in the following years.
Many potential candidates have been discovered, including MnBi, MnAl, and FeNi magnets~\cite{hasegawa_challenges_2021,skomski_magnetic_2016}.
Currently, rare-earths prices tend towards levels similar to those during the crisis period, encouraging further efforts towards developing efficient rare-earth-free permanent magnets.


One of the good alternatives are transition-metal-based (TM-based) magnets.
Fe\nobreakdash-Co alloys are especially promising in this category.
Burkert \etal{} showed, using density functional theory (DFT) calculations, that in uniaxially strained body-centered tetragonal (bct) disordered iron--cobalt (\feco{}) alloys, giant magnetocrystalline anisotropy energy (MAE) of about 800~\muevat{} (over 10~\mjmqb{}) can be achieved for Co concentration $x$ close to 0.6 and lattice parameter ratio $c/a$ close to 1.22~\cite{burkert_giant_2004}.
Such MAE value is comparable to properties observed for SmCo$_5$, Nd$_2$Fe$_{14}$B, and FePt, while at the same time, the saturation magnetization of Fe\nobreakdash-Co significantly exceeds the values observed for the aforementioned materials.
Afterward, many systems have been synthesized following the epitaxial Bain path~\cite{alippi_strained_1997}, including \feco{}/Pt multilayers~\cite{andersson_perpendicular_2006,warnicke_magnetic_2007,andersson_structure_2006} and deposition of \feco{} on Pd~(001)~\cite{winkelmann_perpendicular_2006,yildiz_volume_2009,yildiz_strong_2009}, Ir~(001)~\cite{yildiz_volume_2009,yildiz_strong_2009}, and Rh~(001) buffers~\cite{yildiz_volume_2009,yildiz_strong_2009,luo_tuning_2007}. 
However, the thin-film experiments showed MAE values lower than those predicted by Burkert \etal{}


Neise~\etal{}~\cite{neise_effect_2011} showed that the discrepancies between the theoretically predicted MAE and the measured values could be attributed to the virtual crystal approximation (VCA) utilized by Burkert~\etal{} 
Using 2 $\times$ 2 $\times$ 2 supercell approach with atoms arrangements modeled according to randomized nearest neighbors patterns, they showed that ordered phases of \feco{} have larger MAE than disordered ones, which was confirmed later by Turek~\etal{}~\cite{turek_magnetic_2012}.
They also proposed the preparation of the \feco{} epitaxial films along the Bain path~\cite{alippi_strained_1997}, which has since been realized by Reichel~\etal{}~\cite{reichel_increased_2014,reichel_lattice_2015,reichel_origin_2017} on the Au$_x$Cu$_{1-x}$ buffer, offering a possibility to tailor the lattice parameter in a wide range~\cite{kauffmann-weiss_bain_2014}.


Turek~\etal{} further improved the theoretical prediction, ascribing again the calculated \textit{versus} experimental MAE difference (of the order of 3 -- 4) to the VCA. 
Utilizing a more sophisticated method of the chemical disorder approximation, namely coherent potential approximation (CPA)~\cite{soven_coherentpotential_1967}, they obtained MAE of much lower and a less sharp maximum of 183~\muevat{} spanning a wider range between about 0.5 and 0.65 Co concentration for $c/a$ $\approx$ 1.22~\cite{turek_magnetic_2012}. 
They also showed that ordering of the \feco{} alloys towards \loz{} phase (derived from B2 CsCl structure elongated along the z-axis) could significantly increase the MAE (by a factor between 2 and 3) to 450~\muevat{} for Fe$_{0.4}$Co$_{0.6}$ and 580~\muevat{} for \loz{} Fe$_{0.5}$Co$_{0.5}$ -- corresponding well with theoretical \ku{} of 520~\muevat{} from Ref.~\cite{neise_effect_2011}. 
Importantly, however, experiments and further calculations have shown that \feco{} bct thin films are prone to a rapid relaxation towards the body-centered cubic (bcc) structure above the critical thickness of about 15 monolayers (about 2~nm)~\cite{kim_origin_2013,reichel_increased_2014}.  


Additions of small interstitial atoms such as B, C, and N were proposed to stabilize the necessary tetragonal distortion by the formation of \feco{} martensite phase.
Using special quasirandom structures (SQS) method~\cite{zunger_special_1990} in \fecoc{} supercells, multiple authors obtained a bct structure with $c/a$ lattice parameters ratio as high as 1.12 -- 1.17~\cite{delczeg-czirjak_stabilization_2014,reichel_soft_2015}. 
Several experimentally obtained systems have confirmed these predictions~\cite{reichel_increased_2014,reichel_lattice_2015,reichel_origin_2017,giannopoulos_large_2015,giannopoulos_coherently_2018}, although there is still plenty of room for further improvements.
Two above-mentioned MAE enhancement methods, namely (i) strain induced by a lattice mismatch between two epitaxially grown layers and (ii) spontaneous lattice distortion due to impurities, are summarized in the recent review by Hasegawa~\cite{hasegawa_challenges_2021}. 


Steiner~\etal{} performed an \feco{} case study by averaging over completely random structures in a 2~$\times$~2~$\times$~2 supercell~\cite{steiner_calculation_2016}. 
They suggested that proper caution has to be placed on the averaging method since CPA and VCA are effective medium methods that do not describe local structure relaxation and reduced symmetry. 
Despite their concerns, they obtained MAE values similar to the CPA results reported previously by Turek~\etal{}~\cite{turek_magnetic_2012}. 
Since then, many articles have focused on a supercell approach applied to selected cases of \feco{} doped with boron~\cite{khan_potential_2014}, carbon~\cite{khan_magnetic_2015}, and nitrogen~\cite{chandran_effect_2007,khan_magnetic_2015,odkhuu_firstprinciples_2019}, mostly regarding either (i) the \loz{} phase derived from B2 (CsCl) structure strained along the z-axis, or (ii) the Fe$_{0.4}$Co$_{0.6}$ disordered alloy.
An interesting new way of elucidating the interactions in (\feco{})$_2$B was proposed by D\"ane~\etal{}
They performed a sampling of the full configuration space of the 12-atom supercell, again using the argument that VCA and CPA do not correctly describe the distribution of possible values of MAE and the influence of chemical neighborhood and local geometry optimization. 
They observed a significant spread of the MAE values with an overall average in good agreement with the experiment. 
They argue that treating a ''true'' disorder is certainly beneficial. 
They also noted that it is necessary to average over sufficiently large supercells, as the supercell size can significantly affect the MAE values obtained~\cite{dane_density_2015}.


The discussion about configuration space analysis is connected with symmetry and ordering in the supercell.
Given the vast data set regarding multiple structures in a single crystal system, analysis of ordering towards specific structures is straightforward to implement; it provides more insight into physical phenomena occurring.
Works on energy states of closely related structures reach the '30s--'60s of the 20$^{th}$ century, including contributions from Bethe, Bragg, Williams, Warren, and Cowley in short-range and long-range order analysis methods of that period~\cite{bragg_effect_1934,bragg_effect_1935,bethe_statistical_1935,cowley_approximate_1950,cowley_shortrange_1965}.
Recently, a notable example of ordering effects analysis closely related to our work includes research on the FeNi ordering towards the L1$_0$ phase performed by Izardar, Ederer, and Si~\cite{izardar_impact_2022,izardar_interplay_2020,si_effect_2022}.


In this work, we aim at adding new involvement into the discussion about magnetism in \fecoc{} by implementing a similar method as used by D\"ane~\etal{}
It gives us the possibility to investigate more deeply the possible values of magnetic properties in the system, including dependencies on ordering and, broadly researched, the impurity first coordination shell occupancy.
We present a complete analysis of all stoichiometric compositions modeled in a 2~$\times$~2~$\times$~2 supercell.
We consider all possible symmetrically inequivalent arrangements of Fe and Co atoms.
The aim of the study is to predict the phase stability and intrinsic magnetic properties for the full range of concentrations of the \fecoc{} system and place it in the frame of works on Fe\nobreakdash-Co, Fe\nobreakdash-Co\nobreakdash-B, Fe\nobreakdash-Co\nobreakdash-N, and Fe\nobreakdash-Co\nobreakdash-C alloys.
To achieve it, we study the full configuration space of the 17-atom representation of the Fe\nobreakdash-Co\nobreakdash-C system and explore this approach to crystallize the most effective method of similar analyses for future applications.
Taking advantage of the opportunity to analyze a significant portion of the configuration space of this alloy, we start with the analysis of the entropic and enthalpic contribution towards the system energy.
Selected computational cell size is currently the practical limit due to the computational cost.

\section{Calculations' details}

\subsection{System preparation}


We used the full-potential local-orbital (FPLO18.00) code~\cite{koepernik_fullpotential_1999,eschrig_chapter_2004} with the generalized gradient approximation (GGA) exchange-correlation functional in the Perdew, Burke, and Ernzerhof (PBE)~\cite{perdew_generalized_1996} parametrization for all calculations. 
The use of FPLO was dictated by, \textit{inter alia}, the inherent implementation of the full-potential approach (i.e., omitting the crystalline potential shape approximation), and the expansion of the extended states in terms of localized atomic-like numerical orbitals basis~\cite{koepernik_fullpotential_1999,eschrig_chapter_2004}.
The full-potential approach is particularly essential for accurately determining a subtle quantity such as MAE.
Another important factor in choosing FPLO is the very high performance of the code, at the expense of the lack of multithreading.
In our approach, scaling multiple single-thread calculations up in an embarrassingly parallel manner is the optimal solution.


\begin{figure*}[ht!]
\centering
	\includegraphics[clip,width=0.85\textwidth]{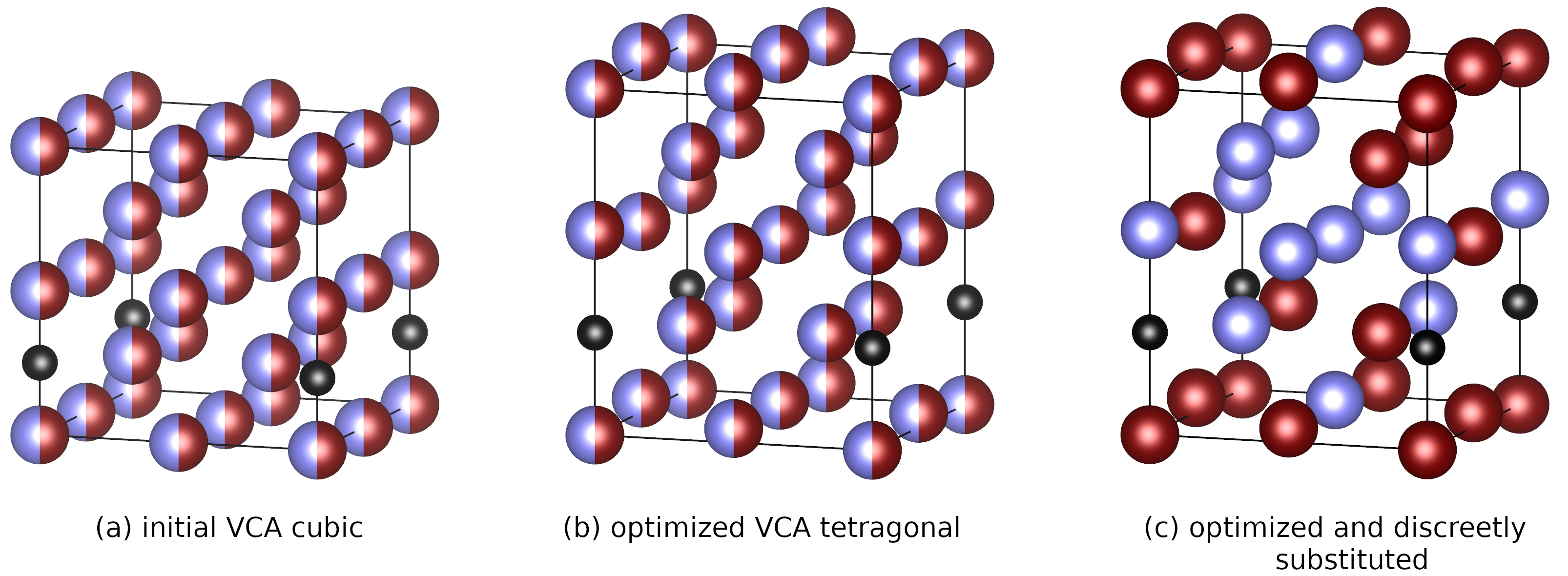}
	\caption{\label{fig-structures}
	Examples of prepared and obtained crystal structures of Fe$_8$Co$_8$C. Initial cubic supercell -- input to virtual crystal approximation (VCA) relaxation (a), structure resultant from VCA geometry optimization (b), and one of the final structures with VCA atoms substituted by Fe and Co atoms (c). Iron, cobalt, and carbon atoms are presented in dark red, light blue, and black, respectively.}
\end{figure*}

Initially, we built a 2~$\times$~2~$\times$~2 supercell of the 2-atom \feco{} body-centered system representation in the $P$4/$mmm$ space group (s.g. 123).
The result is a computational cell containing a total of 16 Fe/Co atoms.
Initial atomic positions were assumed to be perfect (0, 0, 0) and (1/2, 1/2, 1/2) in each unit cell, and a single C atom was introduced as an octahedral interstitial dopant on the (0, 0, 1/4) site in the supercell.
The resultant structure is shown in Fig.~\ref{fig-structures}(a).
Structure visualizations were prepared in VESTA software~\cite{momma_vesta_2008}.
The carbon concentration in the prepared models is about 6 at\% and 1.25 wt\% (1 C atom per 16 TM atoms).
%
%
Initial atomic positions were optimized for Co concentrations equivalent to all stoichiometric cases in the 17-atom supercell (Fe$_{16}$C, Fe$_{15}$CoC, Fe$_{14}$Co$_{2}$C, ..., Co$_{16}$C).
At this stage, we used VCA for the disorder treatment, 6~$\times~$6$~\times~$6 \textbf{k}-point mesh, 10$^{-5}$ density and 10$^{-7}$ Ha ($\sim$2.72 10$^{-5}$~eV) energy convergence criteria and 10$^{-3}$~eV~\AA$^{-1}$ force tolerance for initial optimization.
Cell volume and $c/a$ optimization were performed based on a third-order surface fit to energy \textit{versus} computational cell volume in the 160 -- 208~\AA{}$^3$ range, incremented by 1~\AA{}$^3$ and $c/a$ ratios in the 1.05 -- 1.16 range, incremented by 0.01.
Uniaxial elongation of the cell was assumed after Reichel~\etal{}~\cite{reichel_soft_2015}.
The preparation of the VCA system ended with a full optimization of atomic positions for the minimum of the mentioned fit. 
We used a scalar-relativistic approach with the same parameters as before. 
An exemplary resultant structure for the Fe$_8$Co$_8$C system is shown in Fig.~\ref{fig-structures}(b).


\begin{table}
    \setlength{\tabcolsep}{3.6pt}
    \caption{\label{tab-structures}Number of possible structures in 2 $\times$ 2 $\times$ 2 supercell of AB binary alloy in different representations.}
    \centering
    \begin{tabular}{c|c|c|c|c}
    \hline
    \hline
    Minority atoms                  & bcc       & bct       & bct symmetry   & all       \\
    in AB 2 $\times$ 2 $\times$ 2   & symmetry  & symmetry  & (octahedrally  & possible  \\
    supercell                       & unique    & unique    & doped) unique  &           \\
    \hline
        0   &   1   &   1   &   1   &   1       \\
        1   &   1   &   1   &   5   &   16      \\
        2   &   4   &   6   &   24  &   120     \\
        3   &   6   &   10  &   69  &   560     \\
        4   &   15  &   30  &   174 &   1820    \\
        5   &   17  &   39  &   330 &   4368    \\
        6   &   24  &   67  &   526 &   8008    \\
        7   &   27  &   77  &   694 &   11440   \\
        8   &   32  &   84  &   748 &   12870   \\
    \hline
    \hline
    \end{tabular}
\end{table}

In the final step of structures' preparation, atomic sites were populated with all possible discrete, stoichiometric, geometrically inequivalent Fe/Co occupations.
The equivalency was determined based on the initial, perfect body-centered tetragonal geometry with a single octahedral dopant. 
4 394 unique combinations were obtained out of 65 536 total combinations without repetitions, including 748 unique combinations out of 12 870 for the Fe$_8$Co$_8$C case alone. 
The detailed presentation of the number of unique atomic arrangements in a 16-atom 2 $\times$ 2 $\times$ 2 supercell on a bcc, bct, and octahedrally doped bct lattice are presented in Table~\ref{tab-structures}.
The criterion of identity between the combinations was the equity of all interatomic distances between all atom types, i.e., Fe--Fe, Co--Co, Fe--Co, Fe--C, and Co--C in the initial, perfect supercell. 
It can be proven that it is unambiguous and directly couples each combination with the distribution of minority atoms in the supercell, such as the short-range ordering parameter described later.
This approach provided us with a relatively simple method for preliminary analysis.
Electron density was then converged in the scalar-relativistic mode, using 9~$\times$~9~$\times$~9 \textbf{k}-points over the entire Brillouin zone, following five additional force optimization steps for every structure to prevent numerical artifacts. 
For this step of the calculations, convergence criteria were set at 10$^{-6}$ density and 10$^{-8}$~Ha ($\sim$2.72 10$^{-6}$~eV).
One of the final Fe$_8$Co$_8$C structures is presented in Fig.~\ref{fig-structures}(c).


In the end, we performed a single-step of fully relativistic calculations with magnetization direction aligned in two orthogonal directions, [1~0~0] and [0~0~1], over a charge density self-consistently converged in the scalar-relativistic approach~\cite{daalderop_magnetocrystalline_1991}, a method proven previously to be both accurate and effective~\cite{werwinski_initio_2017,musial_structural_2022}.
Based on the resultant charge density and systems' energies, we derived relevant magnetic and structural parameters per configuration.
Those include magnetocrystalline anisotropy energies (MAE), mixing enthalpies ($\Delta H_{\rm mix}$), magnetic hardness parameter ($\kappa$), Bethe short-range order parameter ($\sigma$), Warren-Cowley short-range order parameter ($\alpha^{XY}$) for first coordination shell, and long-range ordering parameter towards B2 phase ($S$).
Specific equations and methods relevant to detailed parts of the presented work are introduced further alongside the results.

\subsection{Assumptions and ensemble averaging methods}


We estimate our MAE results for each data point to be within 15\% relative error due to relatively low \textbf{k}-point mesh.
Obviously, obtaining accuracy within 1\% for each considered structure would be highly valuable.
However, raising the accuracy would greatly increase the computational cost beyond current capabilities.
Obtained system energies and the mixing enthalpies are much more accurate.
Bound by this limitation, we focus on qualitative trends and averages in more subtle values, such as MAE.
We assume the error imposed by the low \textbf{k}-point mesh for each data point is random and non-cumulative.


We utilize thermal averaging after D\"ane~\etal{}~\cite{dane_density_2015} to include influence of non-optimal ground level energy states:

\begin{equation}
	\label{eqn-averaging}
	{\rm MAE}(T) \equiv \frac{\sum\limits_{\nu}\left[{\rm MAE}_{\nu} \cdot exp(-E_{\nu}/{\rm k_B}T) \cdot n_\nu \right]}{\sum\limits_{\nu}\left[ exp(-E_{\nu}/{\rm k_B}T) \cdot n_\nu \right]}
\end{equation}

\noindent where $E_\nu$ denotes the total energy of a unique atomic arrangement combination $\nu$, MAE$_{\nu}$ represents its magnetocrystalline anisotropy energy, and $n_{\nu}$ is the number of geometrically equivalent configurations. 


An important part of the discussion is whether the averaging assumed in Eq.~\ref{eqn-averaging} is proper.
Foremost, we acknowledge the fact that at room temperature, a vast part of the system does not occupy the ground state, which is calculated in plain DFT.
It results, e.g., in the real magnetic moments being lower than predicted.
A fact more important for us is that Eq.~\ref{eqn-averaging} does not count factors such as the energy barrier height between atom arrangements in the cell.
In fact, if the energy barrier is high enough, simple arithmetic averaging should be more appropriate.
The height of the energy barrier between the conformations could be obtained by, for example, the nudged elastic band (NEB) method~\cite{henkelman_climbing_2000,henkelman_improved_2000}. 
However, it would be computationally not yet feasible to obtain heights of all possible transitions~\cite{rong_efficient_2016}.
Obtaining at least a few values of the barriers in the near future could be beneficial. 
The solution is, however, not compatible with our methods.
Less accurate but less costly linear scaling DFT methods could be utilized to obtain rough values of the barriers.
Moreover, this thermodynamic approach results in the configurations' statistical distribution corresponding to slow cooling. 
We can further assume that despite the obtained results do not rely solely on the most optimal atomic arrangements, the lowest energy structures vastly contribute to the overall MAE.
Overall, Eq.~\ref{eqn-averaging} certainly works for situations corresponding to slow cooling of the alloy.
Hence, it is another assumption in our work that applies to thermal averages.

Apart from the assumptions, an important factor to note is the notation we use to describe various C impurity nearest-neighbor patterns.
Those designations (Fe\nobreakdash-C\nobreakdash-Fe, Co\nobreakdash-C\nobreakdash-Co, and especially Fe\nobreakdash-C\nobreakdash-Co) should not be mistaken with the common Fe\nobreakdash-Co\nobreakdash-C system designation, which we also utilize in this work.

\section{Results and Discussion}

\subsection{Structural properties}



\begin{figure}[ht]
\centering
	\includegraphics[clip,width=\columnwidth]{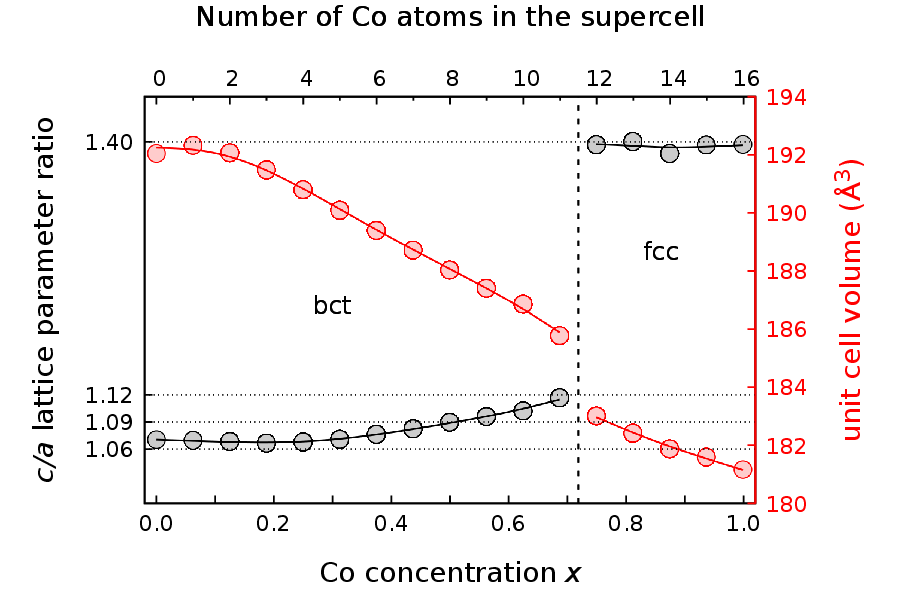}
	\caption{\label{fig-ctoa-v-vs-x}
	Dependency of lattice parameters ($c/a$) ratio (black) and unit cell volume (red) on Co concentration $x$ in \fecoc{} system, calculated with FPLO18 in virtual crystal approximation (VCA) with PBE exchange-correlation potential. The dashed line denotes the structural phase transition between body-centered tetragonal (bct) and face-centered cubic (fcc) structures.}
\end{figure}


We will first discuss the structural parameters of the alloys under consideration.
During the VCA geometry optimization, we observed a structural phase transition from body-centered tetragonal (bct) to face-centered cubic (fcc) structure, which occurs between 11 and 12 Co atoms in the supercell (between 69 and 75\% Co concentration), see Fig.~\ref{fig-ctoa-v-vs-x}.
It corresponds to the well-known phase transition towards hexagonal close-packed structure for high Co concentration in \feco{}. 
The fcc structure is the closest to the hcp structure we can obtain under the assumed constraints.
Although unstable at the standard conditions, the fcc structure for pure Co has been obtained in the high-pressure regime by Yoo~\etal{}~\cite{yoo_new_2000}.


Unit cell volume decreases monotonically with Co concentration after a weak peak for a single Co atom in the supercell, with a significant drop with the transition from bct to fcc structure.
Distinct maximum in unit cell volume has been argued by Pauling and other authors, as brought recently by D\'iaz-Ortiz~\etal{}, to be of the same nature as a peak in magnetization (Slater-Pauling curve)~\cite{pauling_nature_1938,diaz-ortiz_structure_2006}.
The weak maximum we obtained stays in contradiction with the expected, Slater-Pauling-like shape of the curve brought to attention by Prinz~\cite{prinz_stabilization_1985} and successfully reproduced in calculations, e.g., by D\'iaz-Ortiz~\etal{}~\cite{diaz-ortiz_structure_2006} and Steiner~\etal{}~\cite{steiner_calculation_2016}, with a distinct maximum at around 20--30\% Co in \feco{}.
We ascribe this discrepancy to the presence of the dopant atom in the unit cell.
Nevertheless, a noticeable positive deviation from Vegard's law is apparent.
A similar influence of the small interstitial dopant on the structural (and magnetic) parameters of the system has been observed by Chandran~\etal{} for the \fecon{} system~\cite{chandran_effect_2007}.


The exact lattice parameters obtained using the VCA in the bct regime are $a$ ranging from 2.75~\AA{} for Fe$_5$Co$_{11}$C to 2.82~\AA{} for Fe$_{15}$Co$_1$C, and $c$ ranging from 3.01~\AA{} for Fe$_5$Co$_{11}$C to 3.07~\AA{} for Fe$_{12}$Co$_{4}$C.
Resultant optimized volume of the bct systems ranges from 185.8~\AA$^{3}$ for Fe$_5$Co$_{11}$C to 192.3~\AA$^{3}$ for Fe$_{15}$Co$_1$C.
Consistency with Fe$_{16}$C supercell volume obtained by Delczeg-Czirjak~\etal~\cite{delczeg-czirjak_stabilization_2014} in VASP code (about 196 \AA$^3$) is good, as well as comparison to experimental value (about 183~\AA$^{3}$) obtained by Reichel~\etal{} for (Fe$_{0.4}$Co$_{0.6}$)$_{0.98}$C$_{0.02}$~\cite{reichel_increased_2014}. 
The result for equiatomic (Fe$_{0.5}$Co$_{0.5}$)$_{16}$C (188~\AA$^{3}$) is close to values obtained by Khan and Hong in equiatomic (Fe$_{0.5}$Co$_{0.5}$)$_{32}$C (about 187~\AA$^3$)~\cite{khan_magnetic_2015} and (Fe$_{0.5}$Co$_{0.5}$)$_{32}$N (about 188~\AA$^3$)~\cite{khan_potential_2014}.
It is also close to the result by Odkhuu and Hong for (Fe$_{0.5}$Co$_{0.5}$)$_{16}$N (about 190~\AA$^3$)~\cite{odkhuu_firstprinciples_2019}.
Similar values have also been presented for B\nobreakdash-doped \feco{} alloys by Reichel~\etal{}~\cite{reichel_soft_2015}.
This slight overestimation of the transition metal alloy lattice parameter is an expected behavior of the applied PBE exchange-correlation functional.
Diaz-Ort\'iz~\etal{} provided an excellent review of structural parameters, magnetic moments, and stabilities of \feco{} alloys calculated from first principles.
They listed several other results of unit cell volume for \feco{}, ranging from 180 to 190~\AA{}$^3$ per 16-atom cell~\cite{diaz-ortiz_structure_2006}.
Most importantly, Delczeg-Czirjak~\etal{} showed that lattice parameters do not exhibit any significant dependency on the atomic configuration exemplified by the C impurity nearest neighbors~\cite{delczeg-czirjak_stabilization_2014}.
We followed the assumption of not optimizing lattice parameters for every configuration, as it would be too computationally demanding.


Derived lattice parameters lead to the $c/a$ ratio in the bct regime rising from 1.07 in the case of Fe$_{16}$C to 1.12 for Fe$_5$Co$_{11}$C.
It is in agreement with the initial assumption of Burkert \etal{} \cite{burkert_giant_2004} and following theoretical estimations of uniaxial strain induction by interstitial impurities \cite{delczeg-czirjak_stabilization_2014,reichel_soft_2015}.
Reichel~\etal{} presented experimental $c/a$ value of 1.05 for B\nobreakdash-doped Fe$_{0.38}$Co$_{0.62}$, and $c/a$ for (Fe$_{0.4}$Co$_{0.6}$)$_{16}$C equal 1.03--1.04, which is lower than the value of approximately 1.10 close to earlier calculations results present in the literature, and also predicted by us.
They provided several possible reasons for the observed difference in their work~\cite{reichel_soft_2015}.
The phase transition from bct to fcc for \fecoc{} has been also previously reproduced computationally by Delczeg-Czirjak~\etal{} for Co concentration around 65 at\%~\cite{delczeg-czirjak_stabilization_2014}.
Uniaxial strain in the order of a few percent has been numerously shown to lead to reasonable MAE values~\cite{burkert_giant_2004,turek_magnetic_2012,neise_effect_2011,steiner_calculation_2016}, which can be further improved, e.g., by buffer-induced effects in thin-film applications~\cite{reichel_increased_2014,reichel_soft_2015,reichel_origin_2017,rong_efficient_2016}.


\subsection{Mixing entropy}

Considering configuration space analysis, an interesting insight can be provided by investigating the enthalpic and entropic contributions to the system energy.
The entropy of mixing (also called configurational entropy) is a statistical parameter describing the system configuration (or conformation) space quantitatively and, in a basic Boltzmann formulation, takes the form:

\begin{equation}
\Delta S_\mathrm{conf} = -k_\mathrm{B} \ln \omega,
\end{equation}

\noindent where $k_\mathrm{B}$ is Boltzmann constant and $\omega$ is the total number of possible arrangements of the system.
Then, for a finite computational supercell of binary alloy, the atomic contribution to configurational entropy can be calculated from equation~\cite{wang_mixing_2011}:

\begin{equation}
\label{eq_S_conf}
\Delta S_\mathrm{conf} (x, N) = - \frac{1}{N}\,k_\mathrm{B}\,\ln \frac{N!}{[Nx]![N(1-x)]!},
\end{equation}

\noindent where $x$ is the concentration of a selected element in the alloy and $N$ is the number of atomic sites occupied by mixed elements in the supercell. It can be also expressed in the shortened form:

\begin{equation}
\label{eq_S_conf_binomial}
\Delta S_\mathrm{conf} (x, N) = - \frac{1}{N}\,k_\mathrm{B}\,\ln \binom{N}{Nx},
\end{equation}

\noindent where $\binom{n}{k} = \frac{n!}{k!(n-k)!}$ is the binomial coefficient. 
Independently of the exact transcription of the above formula, the equation for the ideal $\Delta S$ value for an infinitely large supercell takes the form:

\begin{multline}
\label{eq_S_conf_ideal}
\Delta S_\mathrm{conf-ideal} (x) = \lim_{N \to \infty}  \Delta \,S_\mathrm{conf} (x, N) =\\
= - k_\mathrm{B} [x \ln x + (1-x) \ln(1-x)].
\end{multline}

For the purposes of our analysis, we consider the configurational entropy of a binary alloy, even though our supercells contain three elements (Fe, Co, and C). 
We motivate this approximation by the C atom fixing in the interstitial position so that only Fe and Co atoms are mixed in common sites.

\begin{figure}[t]
\centering
        \includegraphics[clip, width=\columnwidth]{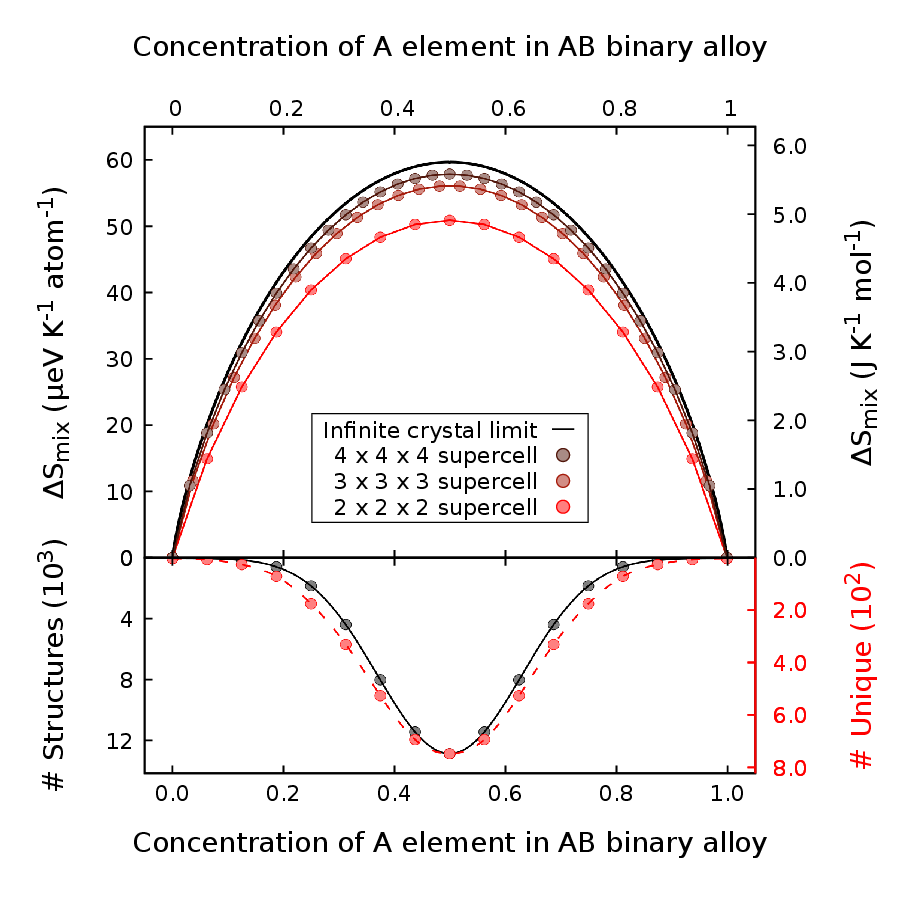}
        \caption{\label{fig-conf-entropy}
        Configurational entropy for binary alloys calculated for several supercells of different sizes and the number of all obtainable structures in 2 $\times$ 2 $\times$ 2 supercells (black). With the dashed red line, we denote the number of symmetrically inequivalent structures in the bct representation with the symmetry broken by an octahedral impurity.
        }
\end{figure}

Figure~\ref{fig-conf-entropy} shows the determination of the Eqs.~\ref{eq_S_conf} and \ref{eq_S_conf_ideal} for several selected supercell sizes and the full range of concentrations, as well as total and reduced number of configurations for the 2 $\times$ 2 $\times$ 2 supercell with the symmetry reduced by a single octahedral interstitial dopant.
The highest configurational entropy value occurs at the equilibrium concentration ($x = 0.5$), and a $2 \times 2 \times 2$ supercell, such as the one we used in this study, leads to an entropy underestimation of about 15\% compared to the ideal result.
The error due to the finite size of supercells decreases quite quickly as their size increases.
The obtained $\Delta S(x)$ relationship can be compared with the previously determined mixing enthalpy values for (Fe,Co)$_{16}$C alloys.
As we can read from Fig.~\ref{fig-conf-entropy}, the highest value of T$\Delta S$ at 1000~K for equiatomic concentration would be equal to 60~meV\,atom$^{-1}$, which in absolute terms is less than the highest value of mixing enthalpy calculated as equal to about -100~meV, for Fe$_{8}$Co$_{8}$C composition.
This means that below 1000~K, there is a high probability that ordered Fe-Co intermetallic compounds will be thermodynamically more favored than solid solutions~\cite{evans_visualizing_2021}.
The estimated gain from considering structure degeneracy was described in the previous sections.

\subsection{Mixing enthalpy and basic magnetic properties \textit{versus} Co concentration}


\begin{figure}[t]
\centering
	\includegraphics[trim=4 6 17 5,clip,width=\columnwidth]{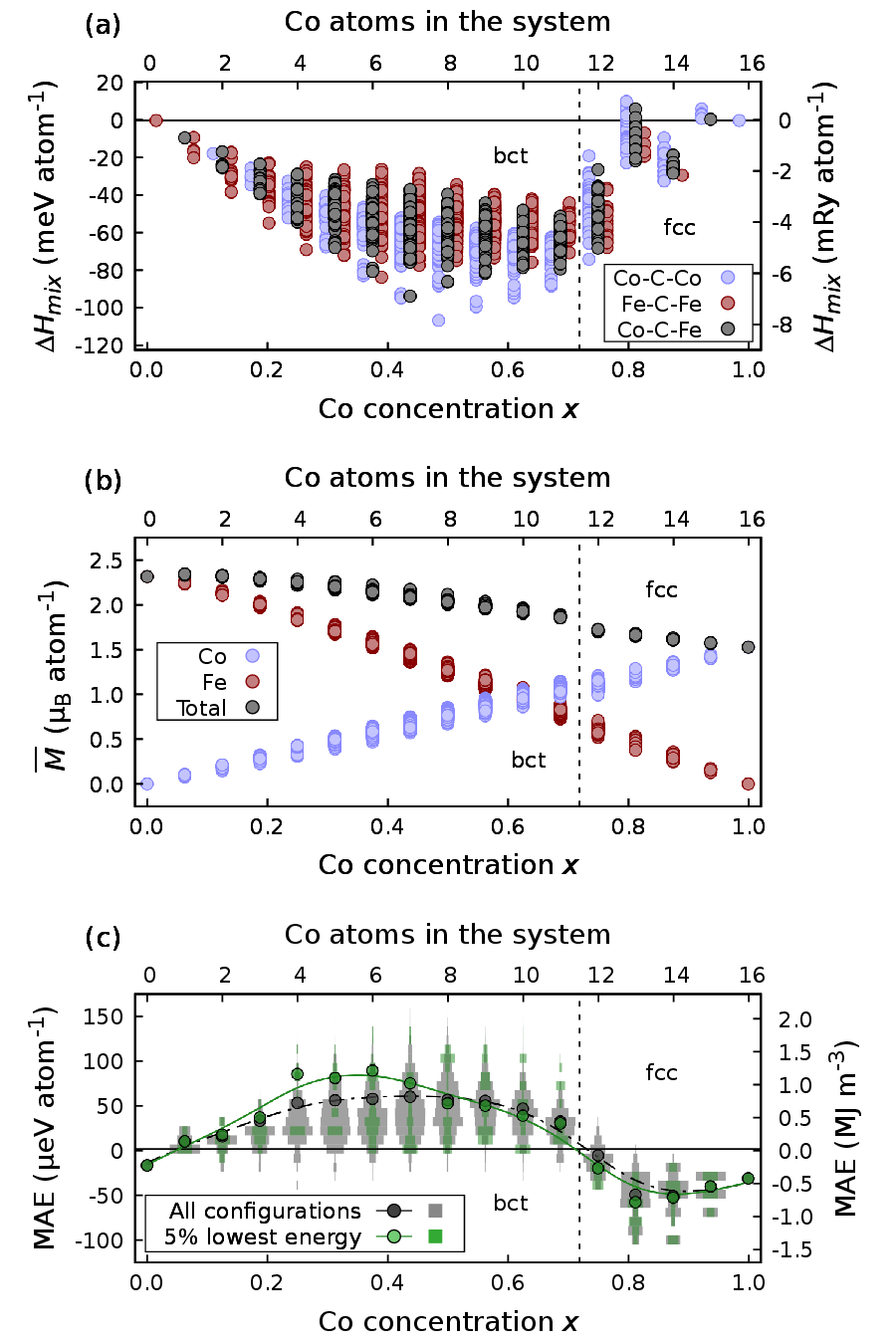}
	\caption{\label{fig-mae-mom-enthalpy-vs-x}
	Mixing enthalpy ($\Delta H_{\rm mix}$), average spin magnetic moments per transition metal atom ($\overline{M}$), and magnetocrystalline anisotropy energy (MAE) per transition metal atom \textit{versus} Co concentration $x$ in \fecoc{} system, as calculated using FPLO18 and PBE exchange-correlation potential for all nonequivalent Fe/Co site occupancies in a 2~$\times$~2~$\times$~2 supercell. 
	In panel (a), the light blue and dark red colors represent systems with, respectively, two Co and two Fe atoms neighboring the C impurity. Dark grey represents systems with one Fe and one Co atom neighboring the dopant.
        For readability, the plotted points are slightly shifted for Co-C-Co and Fe-C-Co configurations.
	The light blue and dark red colors in the (b) panel represent the average contribution of Co and Fe magnetic moments, respectively. 
	Dark gray is the sum of both. 
	Gray histogram (c) represents the aggregation of all results, while the green one represents 5\% of the most energetically favorable atomic arrangements. 
	The circles represent respective average values, and the lines are averaged splines to guide the eye. 
        Vertical dashed lines indicate the structural phase transition between bct and fcc structures.
	}
\end{figure}


A basic energetic parameter describing the system is the mixing enthalpy.
It provides information about the tendency towards the formation of respective structures instead of separation into their constituent phases (in this case, pure Fe- and Co-based phases).
For each structure, we calculated mixing enthalpy $\Delta H_{\rm mix}$ between bct Fe$_{16}$C and fcc Co$_{16}$C using equation analogous to the one used by D\'iaz-Ortiz~\etal{}, for convenient comparison with their results~\cite{diaz-ortiz_structure_2006}:

\begin{equation}
	\Delta H_{\rm mix}(x) = E_{(\rm Fe_{1-x}Co_x)_{16}C} - xE_{\rm Co_{16}C} - (1 - x)E_{\rm Fe_{16}C},
\end{equation}

\noindent as it, in fact, is the same quantity they calculated for ordered \feco{} structures in 2 $\times$ 2 $\times$ 2 supercells. 
The results, presented in Fig.~\ref{fig-mae-mom-enthalpy-vs-x}(a), correspond well with the aforementioned data for \feco{}.
The absolute values of $\Delta H_{\rm mix}$ (up to 8 mRy atom$^{-1}$) are only slightly lower in comparison with up to 9 mRy atom$^{-1}$ calculated by D\'iaz-Ortiz~\etal{}~\cite{diaz-ortiz_structure_2006}.
It indicates the stability of both disordered and ordered \fecoc{} alloys with a minor structure destabilization by the dopant.
Overall, the magnitude of mixing enthalpies suggests good mixing potential, comparable to both TM alloys and steels.
Moreover, the shape of the curve suggests the stability of each of the structures relative to neighboring ones, up to 11 Co atoms in the system, or up to the calculated bct-fcc transition.
Furthermore, a slight asymmetry in the dependence of mixing enthalpy on $x$ can be observed.
On average, the systems closer to the Co-side have lower energies, especially for Co\nobreakdash-C\nobreakdash-Co systems.
However, the absolute minimum for Co\nobreakdash-C\nobreakdash-Co systems occurs for Fe$_8$Co$_8$C.
For Fe\nobreakdash-C\nobreakdash-Co, and especially Fe\nobreakdash-C\nobreakdash-Fe systems, the minimum is moved to the left.
The effect of ordering on the mixing enthalpy will be discussed in the following sections.
On average, for the region around the equiatomic Fe$_8$Co$_8$C, the energy of the systems with C impurity neighbored by two Co atoms is lower compared to systems with the C atom adjacent to two Fe atoms or one Fe and one Co atom.
This is consistent with the observation by Delczeg-Czirjak~\etal{}~\cite{delczeg-czirjak_stabilization_2014} that the energy of Fe\nobreakdash-Co\nobreakdash-C systems depends mainly on the direct chemical neighborhood of the impurity atom, with a preference towards Co\nobreakdash-C\nobreakdash-Co nearest neighbors sequence.
A similar effect has been calculated by Chandran~\etal{} for N\nobreakdash-doped Fe and \feco{}~\cite{chandran_effect_2007}.
Such behavior contradicts the negligence of the direct chemical neighborhood of the impurity atom in earlier works of Khan and Hong~\cite{khan_potential_2014,khan_magnetic_2015,khan_temperature_2019}.
However, we will try to show that despite notable influence on exact quantitative results, negligence of the direct C neighborhood does not alter the qualitative trends in the \fecoc{} system and possibly in other interstitially doped \feco{} systems.
Surprisingly, we can observe a tendency towards the energetic preference of systems containing Fe\nobreakdash-C\nobreakdash-Fe nearest neighbors sequence for low Co concentrations. 
The rapid increase in mixing enthalpy for Co-rich systems is consistent with mixing enthalpies calculated by D\'iaz-Ortiz~\etal{} and instability of Co-rich bct alloys observed in experiments~\cite{diaz-ortiz_structure_2006}.
Meanwhile, in the bct range, the maximum drop in $\Delta H_{\rm mix}$ for any single structure is by a factor of two, and there are no positive-enthalpy (unstable) structures.
It indicates that C doping does not destabilize the system, regardless of the atomic configuration.


In Fig.~\ref{fig-mae-mom-enthalpy-vs-x}(b), we see a decrease in average spin magnetic moments per TM atom with increasing Co concentration.
The average magnetic moment on an Fe atom in Fe$_{16}$C is 2.38~\mub{}, and the average magnetic moment on a Co atom in Co$_{16}$C is 1.53~\mub{}.
There is a positive deviation from a linear change with $x$, similar to the Slater-Pauling-like characteristics of unit cell volume \textit{versus} $x$ dependency.
As seen in partial Fe and Co contributions to the average spin magnetic moment, this deviation from a linear trend stems from the Fe contribution.
The partial contribution from Co magnetic moments increases linearly.
However, as opposed to pure \feco{} results reported by Bardos~\cite{bardos_mean_1969}, we do not observe a characteristic, sharp maximum related to Slater-Pauling behavior.
There is a considerably low deviation in average Fe, Co, and total TM magnetic moments across different configurations.
The structural phase transition, between 11 and 12 Co atoms, affects magnetic moments on both Fe and Co atoms, but the change from 2.55 to 2.48~\mub{} in the average spin magnetic moment on Fe and from 1.56 to 1.46~\mub{} in the average spin magnetic moment on Co, is minimal.

Giannopoulous~\etal{} found magnetization in thin films of (Fe$_{0.45}$Co$_{0.55}$)-C with 20 at\%{} C to be in range of 1600 emu~cc$^{-1}$~\cite{giannopoulos_structural_2015}, which translates to about 2.05~\mubat{}. 
In the literature review performed by Diaz-Ort\'iz~\etal{}, as well as in their own results, we can find average magnetic moments in bcc Fe and bcc Co ranging from 2.13 to 2.35~\mub{} on Fe atoms and from 1.53 to 1.77~\mub{} on Co atoms.
Their MBPP/PBE (mixed-basis pseudopotential code) calculations for ordered Fe\nobreakdash-Co phases yield a total magnetic moment of 2.36~\mubat{} for Fe$_3$Co D0$_3$ phase, 2.29~\mubat{} for Fe\nobreakdash-Co B2 phase, and 2.00~\mubat{} for FeCo$_3$ D0$_3$ phase~\cite{diaz-ortiz_structure_2006}.
Similarly, Chandran~\etal{} reported from VASP/GGA that Fe bcc has a magnetic moment of 2.22~\mubat{}, and Co bcc has a magnetic moment of 1.59~\mubat{}, not counting for the orbital moment contribution, which for both systems should be around 0.10--0.15~\mubat{}~\cite{chandran_effect_2007}.
For C\nobreakdash-doped systems, Delczeg-Czirjak~\etal{} found in SPR-KKR/PBE (spin-polarized relativistic Korringa-Kohn-Rostoker) with CPA that the average magnetic moment drops from 2.2~\mubat{} in systems with the composition close to Fe$_{0.4}$Co$_{0.6}$ to around 1.8~\mubat{} in systems with the compositions close to (Fe$_{0.4}$Co$_{0.6}$)$_{16}$C~\cite{delczeg-czirjak_stabilization_2014}.


Possible giant MAE values are the property that initially brought attention to the \feco{} system.
Hence, MAE is among the first characteristics of the system to consider.
We calculated MAE according to the formula: 

\begin{equation}\label{eqn-MAE}
	\mathrm{MAE} = E_{100} - E_{001},
\end{equation}

\noindent where $E_{100}$ and $E_{001}$ denote the system's energies in the [1~0~0] and [0~0~1] magnetization axis directions (hard and easy axis in the bct structure, respectively).

Figure~\ref{fig-mae-mom-enthalpy-vs-x}(c)~presents MAE \textit{versus} $x$ for all configurations, as well as thermodynamical averages according to Eq.~\ref{eqn-averaging} and assuming $T$~=~300~K for each Co concentration.
We provide an approximate MAE scale~\footnote{To provide an approximate scale in \mjmqb{}, we assume a uniform, average cell volume of 189~\AA$^3$ across all Co concentrations and TM atoms configurations. 
The approximate scale can yield values with uncertainty up to 1.5\% for the bct region} in \mjmqb{}.
Vertical histograms are scaled to fit the width between points and represent the data spread (i.e. represent the probability distribution).
There is apparently an unimodal distribution of all MAE results for the whole $x$ range among all configurations.
A bimodal distribution can be observed in the results obtained for the lowest energy configurations, with MAE values being either very high or near zero.
We observe that MAE varies hugely between configurations, with the absolute maximum for 7 Co atoms in the 16 TM-atom supercell.
With more than 11 Co atoms in the system, we observe a rapid decrease and change in the sign of MAE, associated with the phase transition.
The high difference in MAE between individual configurations is consistent with similar results for ordering towards L1$_0$ phase in equiatomic FeNi obtained by Izardar, Ederer, and Si.
Though we focus on qualitative trends with low convergence criteria for each data point, they conducted a full convergence for several dozen structures~\cite{izardar_impact_2022,si_effect_2022}. 


\begin{figure}[t]
\centering
	\includegraphics[trim=20 3 20 5,clip,width=\columnwidth]{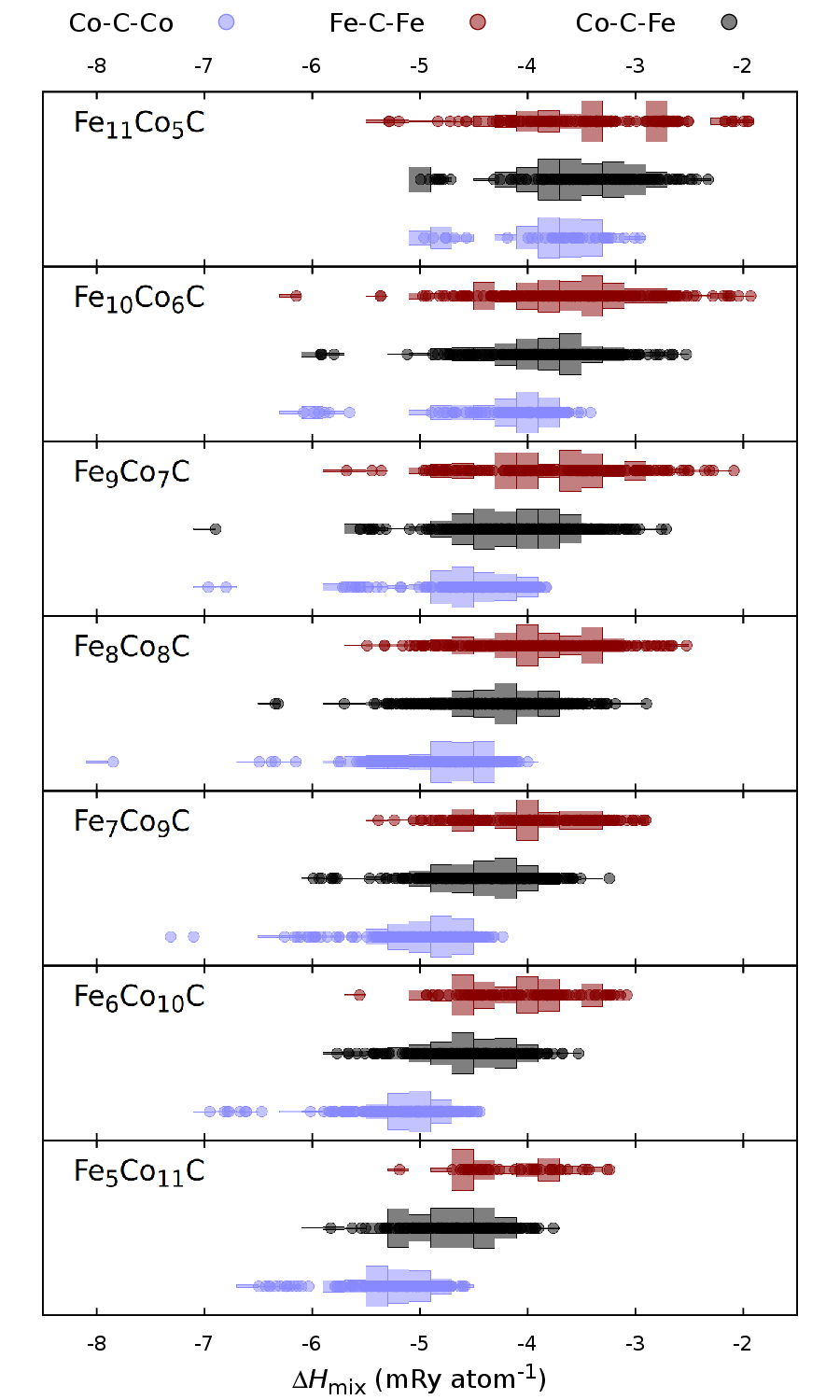}
	\caption{\label{fig-enthalpy-vs-x-histograms}
	    Mixing enthalpies probability distribution across all configurations for intermediate Co contents in the 2 $\times$ 2 $\times$ 2 supercell of \fecoc{}. 
    The light blue and dark red colors represent systems with, respectively, two Co and two Fe atoms neighboring the C impurity. 
        Dark grey represents systems with one Fe and one Co atom neighboring the dopant. 
        Points along the histograms present the distribution of geometrically inequivalent structures and are horizontal representations of data presented in vertical series on Fig.~\ref{fig-mae-mom-enthalpy-vs-x}(a).}
\end{figure}

Closer inspection of Fig.~\ref{fig-mae-mom-enthalpy-vs-x}(c), especially the lowest-energy-configurations-resolved thermal averages (green points connected with the solid green line), leads to observation that the averages lie much higher than the most probable MAE in terms of the presented histograms, especially for Fe$_{12}$Co$_{4}$C, Fe$_{11}$Co$_{5}$C, and Fe$_{10}$Co$_{6}$C.
In the case of those concentrations, considering only 5\% of the most energetically favorable structures leads to the overestimation of the MAE by almost a factor of two.
The conclusion is twofold.
Firstly, the overestimation by a factor of two is better than the overestimation by a factor of four, known for the VCA method and \feco{} system~\cite{burkert_giant_2004,turek_magnetic_2012}.
It is, however, still significant, and the care should be ensured to include enough samples in the statistics.
We took a certain fixed percentage of the lowest energy unique configurations, with a minimum of 10 (as a minimal statistically significant amount), which leads to a substantial influence of entropic effects.
This approach is open for future improvements in the relation between sampled configurations and the configurational entropy of the system, but such enhancements are beyond the scope of this work.
Secondly, the number of ways in which certain configurations can be achieved -- configuration degeneracy due to the resultant structure symmetry -- cannot be the only culprit beyond the discrepancies between the full and partial configuration space sampling.

To investigate the discrepancy between full and reduced statistics, we plotted structure occurrence probability distribution for different mixing enthalpies, and present it in Fig.~\ref{fig-enthalpy-vs-x-histograms}. 
Since the formation enthalpy can be derived in a similar way, with a reference point leading to a shift in values, the distribution is representative of this quantity, too.
An expected feature of the plot would be following the Gaussian distribution, with the vast majority of the systems possessing medium mixing enthalpy, and only a few of really preferable (or unpreferable) structures, with possible addition of ordering effects, leading to a slight change in the curve shape.
Clearly, that is not the case.
Though we can observe such features for the Fe-C-Co structures, and to some extent Co-C-Co structures, the distribution for the Fe-C-Fe structures is definitely not unimodal.
In the case of Fe$_{11}$Co$_5$C or Fe$_5$Co$_{11}$C one could argue that some discrepancies may stem from the relatively low number of samples, but we remind that we got 748 unique structures for the Fe$_8$Co$_8$C system, out of which over 200 contain Fe-C-Fe nearest neighbor sequence.
Each of those configurations can be realized in multiple ways, which were included in the histograms.
We ascribe this behavior to various ordering effects influencing the systems' energy.
Nevertheless, the most energetically preferable system can have much lower mixing enthalpy (enthalpy of formation\footnote{In the bct region those two quantities are equivalent.}) than the vast majority of the structures.
It leads to its weight in Boltzmann averaging being a few orders of magnitude higher, though the number of systems with the mixing (formation) enthalpy closer to the most probable one can outweigh the contribution from the most energetically preferable configuration.
Hence, regardless of the undoubtful usefulness of effective medium-based methods such as VCA and CPA, taking into account the effects of the system's ordering can be necessary for the proper description of the sensitive electronic properties.

\begin{figure*}[ht]
\centering
	\includegraphics[clip,width=\textwidth]{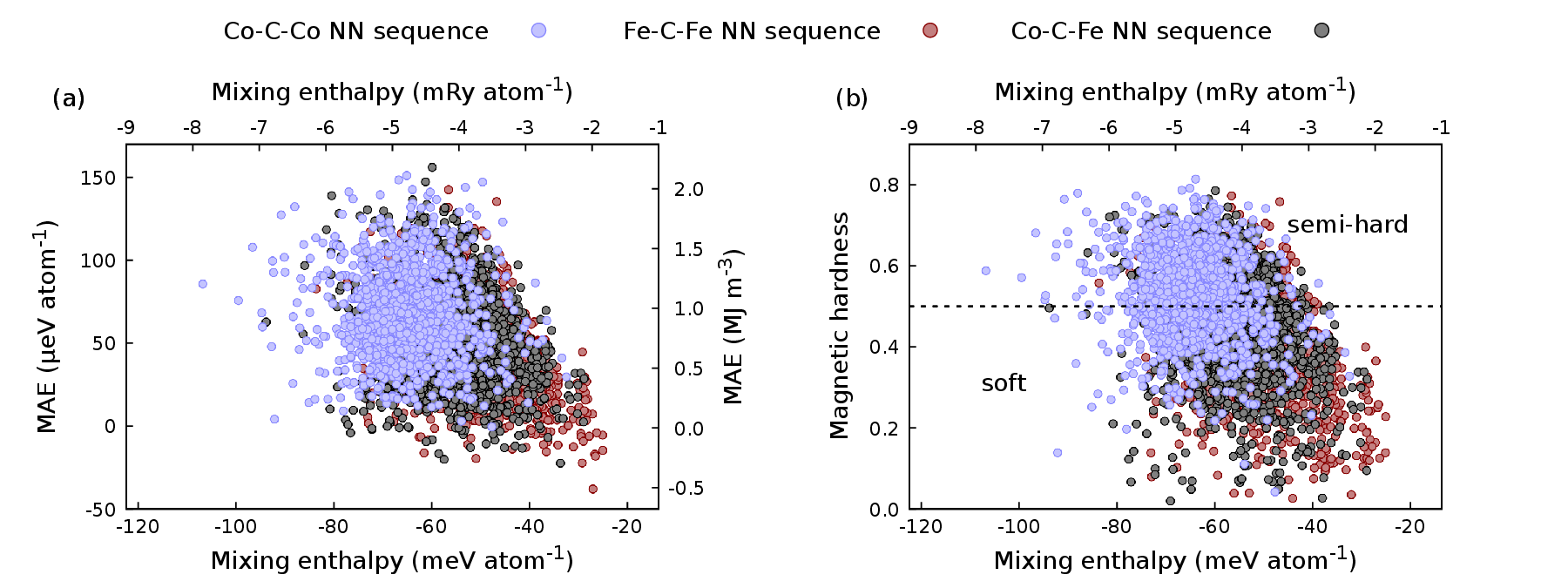}
	\caption{\label{fig-mae-hardness-vs-enthalpy}
	Magnetocrystalline anisotropy energy (MAE) (a), and magnetic hardness (b) \textit{versus} Fe$_{16}$C and Co$_{16}$C mixing enthalpy in the \fecoc{} system (from 3 to 11 Co atoms in the supercell). The light blue color denotes systems with two Co atoms neighboring the C dopant, the dark red color denotes systems with two Fe atoms neighboring the impurity, and the black color denotes systems with the C atom neighbored by one Fe and one Co atom. In panel (b), the dashed line for hardness equal to 0.5 indicates a semi-hard magnetic materials threshold. The results were obtained in FPLO18 with the PBE exchange-correlation potential.}
\end{figure*}

Focusing on qualitative trends, in Fig.~\ref{fig-mae-mom-enthalpy-vs-x}(c), we see a broad maximum for $x$~$\simeq$~0.25~--~0.75.
According to Eq.~\ref{eqn-averaging}, we obtained an average MAE of 0.87~\mjmqb{} for Fe$_8$Co$_8$C. 
MAE decreases by around 20\% between $x$ = 0.5 and $x$ $\simeq$ 0.3.
%
It is in contrast to a rapid drop in MAE for low Co concentrations reported by Delczeg-Czirjak~\etal{} (65\% drop between $x$~=~0.6 and $x$~=~0.3).
We obtained nearly the same MAE values for $x$~$\simeq$~0.6 and $x$~$\simeq$ 0.3.
Intriguingly, we observe several configurations with relatively high MAE values for Co concentration as low as 0.25.
Our findings of notable, positive MAE for low Co concentrations contradict earlier results obtained with effective medium methods.
VCA and CPA reported negative MAE for low Co concentrations Fe\nobreakdash-Co alloy, as seen on MAE \textit{versus} $c/a$ \textit{versus} $x$ maps by Burkert~\etal{} and Turek~\etal{}~\cite{burkert_giant_2004,turek_magnetic_2012}.
On the other hand, it is consistent with the findings of Steiner~\etal{}, who reported positive MAE for \feco{} supercells in a much wider Co concentration range, and Wu~\etal{}, who reported high MAE for Fe$_{12}$Co$_{4}$C and Fe$_{11}$Co$_{5}$C~\cite{steiner_calculation_2016,wu_firstprinciples_2008}.
Moreover, we observe a few high-MAE configurations among the 5\%{} most preferable ones, see green histograms in Fig.~\ref{fig-mae-mom-enthalpy-vs-x}(c).
The thermodynamically averaged MAE values over 5\% of the lowest energy configurations overestimate averages of all symmetrically non-equivalent configurations.
It suggests the non-negligible influence of high-energy (and hence low probability) structures stemming from their quantity.

\begin{table*}
\setlength{\tabcolsep}{3.6pt}
\caption{\label{tab-mae}Magnetocrystalline anisotropy energy of selected \fecoc{} systems, resultant from the averaging proposed in Eq.~\ref{eqn-averaging} and calculated in FPLO18 with PBE exchange-correlation functional, compared to selected other computational and experimental results. Reference results were chosen to resemble our setup as closely as possible.}
\begin{tabular}{c|c|c|c|c}
\hline
\hline
Source                                                            & Compound                                       & Method                                & MAE (\muevat{}) & MAE (\mjmqb{})  \\
\hline
This work                                                         & Fe$_{11}$Co$_{5}$C                             & Calc. supercells (FPLO)               & 51              & 0.69            \\
This work                                                         & Fe$_{8}$Co$_{8}$C                              & Calc. supercells (FPLO)               & 64              & 0.87            \\
This work                                                         & Fe$_{5}$Co$_{11}$C                             & Calc. supercells (FPLO)               & 40              & 0.55            \\
Giannopulous \etal{} \cite{giannopoulos_structural_2015}          & (Fe$_{0.45}$Co$_{0.55}$)$_{0.8}$C$_{0.2}$      & Exp. SQUID/EMCD, thin film on Au-Cu   & --              & $\sim$ 0.8      \\
Giannopulous \etal{} \cite{giannopoulos_large_2015}               & Fe$_{0.8}$C$_{0.2}$/Co, Fe/Co$_{0.8}$C$_{0.2}$ & Exp. SQUID, thin film on Au-Cu        & --              & $\sim$ 1        \\
Giannopulous \etal{} \cite{giannopoulos_coherently_2018}          & (Fe$_{0.45}$Co$_{0.55}$)$_{0.9}$C$_{0.1}$      & Exp. FMR, FeCo-C 3nm/Au-Cu 1nm layers & --              & up to 0.4       \\
Reichel \etal{} \cite{reichel_lattice_2015}                       & (Fe$_{0.4}$Co$_{0.6}$)$_{0.98}$C$_{0.02}$      & Exp. VSM, thin films on Au-Cu         & --              & 0.8 $\pm$ 0.15  \\
Reichel \etal{} \cite{reichel_increased_2014}                     & (Fe$_{0.4}$Co$_{0.6}$)$_{0.98}$C$_{0.02}$      & Exp. VSM, thin films on Au-Cu         & --              & 0.44 $\pm$ 0.14 \\
Reichel \etal{} \cite{reichel_increased_2014}                     & (Fe$_{0.4}$Co$_{0.6}$)$_{32}$C                 & Calc. VCA (WIEN2k) / CPA (SPR-KKR)    & --              & 0.51 / 0.224    \\
Reichel \etal{} \cite{reichel_soft_2015}                          & Fe$_{0.38}$Co$_{0.62}$-B                       & Exp. VSM, thin films on Au-Cu         & --              & 0.4--0.55       \\
Reichel \etal{} \cite{reichel_soft_2015}                          & (Fe$_{0.5}$Co$_{0.5}$)$_{16}$B                 & Calc. CPA (SPR-KKR)                   & 46              & 0.62            \\
Reichel \etal{} \cite{reichel_soft_2015}                          & (Fe$_{0.4}$Co$_{0.6}$)$_{16}$B                 & Calc. CPA (SPR-KKR)                   & 52              & 0.69            \\
Reichel \etal{} \cite{reichel_origin_2017}                        & (Fe$_{0.4}$Co$_{0.6}$)$_{0.98}$C/B$_{0.02}$    & Exp. FeCo-C/B 4nm/Au-Cu 4nm layers    & --              & $\sim$ 0.5      \\
Delczeg-Czirjak \etal{} \cite{delczeg-czirjak_stabilization_2014} & (Fe$_{0.4}$Co$_{0.6}$)$_{16}$C                 & Calc. VCA (WIEN2k)                    & 90              & 1.29            \\
Delczeg-Czirjak \etal{} \cite{delczeg-czirjak_stabilization_2014} & (Fe$_{0.4}$Co$_{0.6}$)$_{16}$C                 & Calc. CPA (SPR-KKR)                   & 42              & 0.59            \\
Delczeg-Czirjak \etal{} \cite{delczeg-czirjak_stabilization_2014} & Fe$_{6}$Co$_{10}$C                             & Calc. SQS (VASP)                      & 51 $\pm$ 9      & 0.75 $\pm$ 0.13 \\
Odkhuu and Hong \cite{odkhuu_firstprinciples_2019}                & Fe$_{27}$Co$_{27}$N$_{2}$                      & Calc. ordered supercells (VASP)       & $\sim$ 100      & --              \\
Khan and Hong \cite{khan_magnetic_2015}                           & Fe$_{64}$Co$_{64}$C$_{4}$                      & Calc. ordered supercells (FLAPW+VASP) & 47              & 0.62            \\
Khan and Hong \cite{khan_magnetic_2015}                           & Fe$_{64}$Co$_{64}$N$_{4}$                      & Calc. ordered supercells (FLAPW+VASP) & 42              & 0.58            \\
Khan and Hong \cite{khan_potential_2014}                          & Fe$_{64}$Co$_{64}$N$_{4}$                      & Calc. ordered supercells (FLAPW+VASP) & 59              & 0.8             \\
\hline
\hline
\end{tabular}
\end{table*}

Our quantitative MAE results can be placed in the context of numerous works describing selected atomic configurations in pure Fe\nobreakdash-Co, as well as B-, C-, and N\nobreakdash-doped systems, realized both experimentally and by DFT calculations to date.
For comparison, in Table~\ref{tab-mae}, we present our thermally averaged MAEs along with selected results from the literature.
Data for cross-reference is chosen to resemble our setup ($c/a$ ratio, and TM and dopant concentration) as closely as possible.
At the beginning, we want to point out a few noticeable general features of some of the results we are comparing our results to.
Firstly, some of the samples analyzed in experiments, i.e. references~\cite{giannopoulos_coherently_2018} and~\cite{reichel_origin_2017}, were prepared as layered systems with various content of Au-Cu interlayer buffers.
Hence, the results of those experiments can serve as a lower boundary of technologically possible MAE.
Secondly, in multiple of the cited articles, VCA is proven to overestimate the MAE~\cite{delczeg-czirjak_stabilization_2014,turek_magnetic_2012,reichel_increased_2014}.
Thirdly, highly ordered structures, like the one calculated by Odkhuu and Hong~\cite{odkhuu_firstprinciples_2019}, can present heightened MAE -- as shown by Turek~\etal{}~\cite{turek_magnetic_2012} and as we prove in this work.
Going into the more precise description, Giannopoulos~\etal{} found experimentally \ku{} for C\nobreakdash-doped Fe$_{0.45}$Co$_{0.55}$ thin films to be in order of 0.8~\mjmqb{}~\cite{giannopoulos_structural_2015}, exact same value as obtained by Reichel~\etal{} for (Fe$_{0.4}$Co$_{0.6}$)$_{0.98}$C$_{0.02}$ thin films~\cite{reichel_lattice_2015}.
Reichel~\etal{} have also shown from combined DFT and experimental analysis that the (Fe$_{0.4}$Co$_{0.6}$)$_{32}$C system possesses slightly lower MAE of the order of 0.5~\mjmqb{} and much higher stability for relatively thick films~\cite{reichel_increased_2014}.
They also reported B\nobreakdash-doped \feco{} alloys to behave similarly, with a little higher MAE than C\nobreakdash-doped system~\cite{reichel_soft_2015}.
Odkhuu and Hong provide similar results for \fecon{}~\cite{odkhuu_firstprinciples_2019}.
Delczeg-Czirjak~\etal{} found MAE for Fe$_{6}$Co$_{10}$C to be in the order of 51~\muevat{} or 0.75~\mjmqb{} as calculated in WIEN2k/SQS, higher than their SPR-KKR/CPA calculations (41.6~\muevat{})~\cite{delczeg-czirjak_stabilization_2014}.
For B2 Fe\nobreakdash-Co\nobreakdash-C and Fe\nobreakdash-Co\nobreakdash-N systems, Khan and Hong reported MAE values of 0.65 and 0.58~\mjmqb{}, respectively~\cite{khan_magnetic_2015}.

Overall, our results agree well with previous calculations and experiments wherever direct comparison is possible.
Qualitative trends among major magnetic properties are similar, and quantitative results lie close to previous DFT data.
However, the dataset we provide is vastly greater than anything currently available in the literature.
The method implemented in our work is slightly more computationally expensive than the CPA method while yielding MAE results with a similar accuracy.
On the other hand, it allows us to provide a more detailed analysis of aspects other than the basic magnetic properties.
In particular, it enables us to interpret ordering dependencies, which we will present in the following sections.

\subsection{Magnetocrystalline anisotropy energy and magnetic hardness in relation to the mixing enthalpy}


To systematize the dataset, we first analyze the dependency of MAE on the mixing enthalpy.
This dependency for all configurations is shown in~Fig.~\ref{fig-mae-hardness-vs-enthalpy}(a).
We see an increase of MAE with lowering the system enthalpy, indicating the preference towards high-MAE structures.
There is a significant scatter of values for separate systems around the average.
Systems with the dopant atom neighbored by two Co atoms have noticeably larger MAE and lower mixing enthalpy relative to the systems with Fe\nobreakdash-C\nobreakdash-Fe and Fe\nobreakdash-C\nobreakdash-Co nearest neighbors (NN) sequence. 


To further explore the usefulness of investigated structures, we calculate magnetic hardness.
It is a parameter describing the system resistance towards spontaneous self-demagnetization and can be defined as~\cite{skomski_magnetic_2016}:

\begin{equation}
	\label{eqn-kappa-1}
	\kappa = \sqrt{\frac{K_1}{\mu_0M_{\rm S}^2}},
\end{equation}

\noindent where $K_1$ is the magnetic anisotropy constant, $M_S$ is the saturation magnetization, and $\mu_0$ is the vacuum permeability.
A simple empirical rule is that a permanent magnet candidate needs $\kappa$ greater than 1 to resist self-demagnetization.
$\kappa$ is a useful technical value, as plenty of magnets with relatively low MAE values are manufactured widely due to their high magnetic hardness and low materials cost.

In the case of the Fe\nobreakdash-Co\nobreakdash-C system, numerous experimental realizations showed a possibility of further amendment of the system to at least double its MAE by tuning the $c/a$ ratio, where interstitial doping can be combined with growth on specifically tailored substrates~\cite{reichel_lattice_2015,reichel_soft_2015,delczeg-czirjak_stabilization_2014,giannopoulos_coherently_2018}.
We also previously showed the positive effect of 5$d$ doping of a similar system~\cite{musial_structural_2022}.
Hence, we are interested in promising compositions showing at least semi-hard magnetic properties due to C-doping alone.
Skomski and Coey described systems with $\kappa$ around 0.5 as semi-hard~\cite{skomski_magnetic_2016}.
We mark the $\kappa$ = 0.5 value in~Fig.~\ref{fig-mae-hardness-vs-enthalpy}(b) with dashed lines.
In our estimation, we assume $K_1$ equals MAE, as defined before.
Saturation magnetization is derived from the calculated total magnetic moment and cell volume.
Thus, we can expand Eq.~\ref{eqn-kappa-1} to the form:

\begin{equation}
        \label{eqn-kappa-2}
	\kappa = \sqrt{\frac{E_{100} - E_{001}}{\mu_0 \left[ \frac{\sum_{i} M_{i}}{V} \right]^2}},
\end{equation}

\noindent where $i$ is the atomic site in the computational cell, $M_i$ is the total magnetic moment of the atom occupying site $i$, and $V$ is the computational cell volume.

Figure~\ref{fig-mae-hardness-vs-enthalpy}(b) presents the resultant magnetic hardness \textit{versus} mixing enthalpy relation.
It is similar to the MAE dependency on the mixing enthalpy, presented in~Fig.~\ref{fig-mae-hardness-vs-enthalpy}(a).
The magnetic hardness of many configurations exceeds the conventional limit of 0.5 for semi-hard magnetic materials but does not exceed 0.9, remaining below the limit for hard magnetic materials.
Odkhuu and Hong reported similar values of $\kappa$, ranging from 0.5 to 1 for the \fecon{} system~\cite{odkhuu_firstprinciples_2019}.
From Eq.~\ref{eqn-kappa-2}, we can see that there are two main ways to improve the magnetic hardness of the sample.
We can either improve MAE or reduce magnetic moment.
For permanent magnet applications, we are at the same time interested in as high saturation magnetization as possible.
It implies that improving magnetic anisotropy while maintaining relatively high magnetic moments is of interest.
Alternatively, achieving high magnetic hardness at the cost of magnetic moment can be beneficial in case of sufficient economic advantage.
Relatively negligible changes in total magnetic moment across configurations with the same Co content suggest that, in our case, the MAE changes are a decisive factor in the magnetic hardness variations for different configurations.
Either way, both pathways for MAE improvements are feasible in the Fe\nobreakdash-Co\nobreakdash-C system.

\subsection{Magnetic moments}


\begin{figure*}[ht]
\centering
	\includegraphics[clip,width=\textwidth]{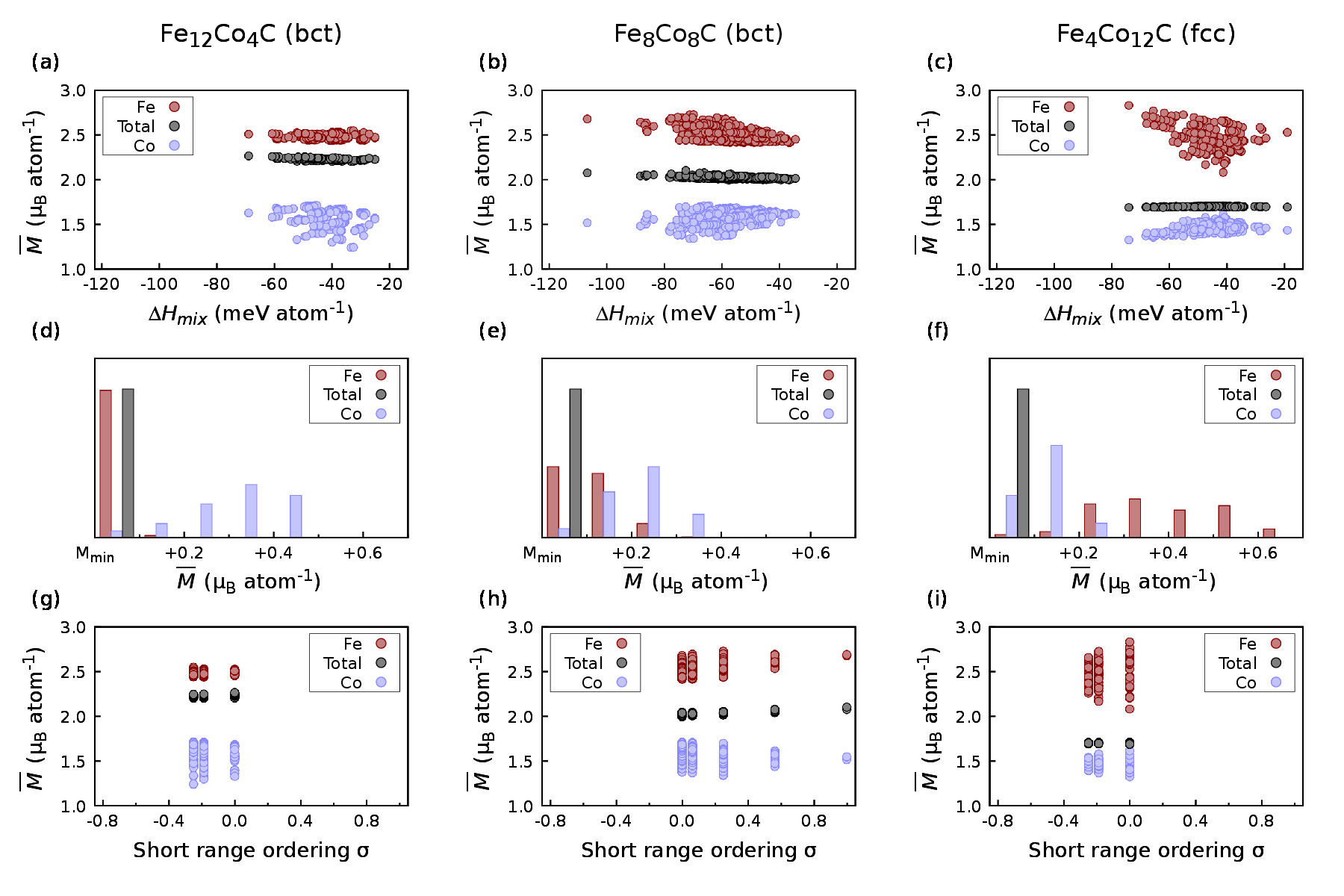}
	\caption{\label{fig-mom-vs-ordering-histogram-enthalpy}
	The average spin magnetic moment per atom \textit{versus} the mixing enthalpy of Fe$_{16}$C and Co$_{16}$C (a--c), the same value presented as a histogram (d--f), and versus Bethe short-range ordering (g--i), obtained for all geometrically inequivalent Fe/Co arrangements in a 2 $\times$ 2 $\times$ 2 supercells of \fecoc{}, calculated with FPLO18 code and PBE exchange-correlation potential. Results are presented for 4, 8, and 12 Co atoms in the supercell. The Fe$_{4}$Co$_{12}$C fcc structure is metastable.}
\end{figure*}


Looking into the dataset, we focus on average magnetic moments per TM atom in the system, along with the spread of the values in different atomic configurations.
Figure~\ref{fig-mom-vs-ordering-histogram-enthalpy} summarizes results for exemplary Co concentrations $x$, 25\%, 50\%, and 75\%.
Presented trends in average Fe, Co, and total spin magnetic moments -- dependencies on mixing enthalpy and short-range ordering, and their distribution -- are representative.
Similar results in the literature are scarce, in contrast to analyses of TM magnetic moments on different impurity atom coordination shells, performed by, e.g., Delczeg-Czirjak~\etal{} and Khan~\etal{}~\cite{delczeg-czirjak_stabilization_2014,khan_magnetic_2015,khan_potential_2014,khan_temperature_2019}.


As presented in~Fig.~\ref{fig-mom-vs-ordering-histogram-enthalpy}(a), for low Co concentration, the low enthalpy configurations are particularly associated with high average magnetic moment on Co atoms.
It can be explained by the preferred Fe\nobreakdash-C\nobreakdash-Fe neighborhood, as the dopant atom tends to lower magnetic moments on neighboring atoms.
Delczeg-Czirjak~\etal{} shown that TM atoms adjacent to the C impurity in the \fecoc{} system have significantly reduced magnetic moments~\cite{delczeg-czirjak_stabilization_2014}.
For intermediate Co content (exemplified by the Fe$_8$Co$_8$C system), there is no significant correlation between the average total magnetic moment and mixing enthalpy neither on Fe nor on Co atoms in the bct range.
For $x$~=~0.75 (in the fcc range), a preference towards higher Fe and lower Co magnetic moments emerges.
We can observe that despite the average total spin magnetic moment on Fe and Co atoms varying considerably between configurations, the average total spin magnetic moment per atom remains almost constant.
Spin magnetic moment on Co atoms remains close to 1.5~\mubat{}, as predicted by linearity in its linear partial contribution to the total average spin magnetic moment in the supercell. 


The trend can be seen more clearly in Fig.~\ref{fig-mom-vs-ordering-histogram-enthalpy}(d--f) where we present histograms of the average Fe, Co, and total magnetic moments in the structures.
In general, the magnetic moment on Fe atoms depends much more on their chemical neighborhood than the magnetic moment on Co atoms.
In~Fig.~\ref{fig-mom-vs-ordering-histogram-enthalpy}(d), we see that on the Fe-rich side of the concentration range, for Fe$_{12}$Co$_4$C, the total magnetic moment in the system, 2.233~\mub{}, remains almost constant across all configurations with a triple standard deviation of 0.03~\mub{}.
A similar trend can be observed for the average Fe magnetic moment (2.48$\pm$0.08~\mubat{}).
However, for average Co magnetic moments (1.57~\mubat{}), we can see that the triple standard deviation is relatively high and equals 0.31~\mubat{}.
On the Co-rich side, for Fe$_4$Co$_{12}$C alloy, see~Fig.~\ref{fig-mom-vs-ordering-histogram-enthalpy}(f), we notice that the total magnetic moment in the system also remains almost constant (1.70$\pm$0.02~\mubat{}). 
Still, we observe a noticeable variation of 0.16~\mubat{} around the average value of Co magnetic moments (1.45~\mubat{}).
However, average magnetic moments on Fe atoms, 2.48~\mubat{}, vary considerably across different configurations, in the range of $\pm$0.42~\mubat{}, which yields almost 34\% relative variability between lowest and highest Fe magnetic moment value.
In~Fig.~\ref{fig-mom-vs-ordering-histogram-enthalpy}(e), presenting results for Fe$_8$Co$_8$C, we observe moderate variation in average magnetic moments on both Fe and Co atoms, in the range of 2.53$\pm$0.20~\mubat{} on Fe, 1.56$\pm$0.22~\mubat on Co, and a total magnetic moment in the system of 2.03$\pm$0.05~\mubat{}.

Again, a major driving factor in the spread of magnetic moments across all structures can be the magnetic moment lowering by the neighboring C impurity, which is most prominent on Co atoms, as presented for numerous Fe\nobreakdash-Co\nobreakdash-based systems by Khan~\etal{}~\cite{khan_magnetic_2015,khan_potential_2014,khan_temperature_2019}.
Moreover, a similar result for N\nobreakdash-doped B2 Fe\nobreakdash-Co was obtained by Chandran~\etal{}. 
They obtained magnetic moments being reduced from 2.78~\mub{} to 2.09~\mub{} between next-nearest and nearest neighbors of the dopant for Fe atoms and from 1.76~\mub{} to 1.12~\mub{} for Co atoms, with magnetic moment fluctuations propagating into next-nearest neighbors~\cite{chandran_effect_2007}.


To explore other factors influencing magnetic moments in the system, we can use a local neighborhood-based order parameter $\sigma$ of Bethe, which can be defined for a binary alloy as~\cite{bethe_statistical_1935}:

\begin{equation}
    \label{eqn-bethe}
	\sigma = p_{\rm AB} - (p_{\rm AA} + p_{\rm BB}) = 2p_{\rm AB} - 1,
\end{equation}

\noindent where $p_{XY}$ denotes the probability of finding an XY nearest neighbor pair.

Though developed for equiatomic systems, $\sigma$ derived from Eq.~\ref{eqn-bethe} also provides useful information for non-equiatomic binary systems, as it depicts changes in the system with increasing content of NN pairs of non-similar atoms.
In that case, $\sigma$ generally takes values between -1 and 1, with positive values indicating preference towards dislike (in our case, Fe--Co) atomic pairs in the structure and negative values indicating preference towards same-atom type pair (Fe--Fe and Co--Co).
However, both minimum and maximum achievable $\sigma$ changes with the system composition and supercell size, $\sigma_{min}$ being in <-1,0> range (likewise atom pair affinity) and $\sigma_{max}$ in <0,1> range (dislike atom pair affinity). 

\begin{figure*}[ht!]
\centering
	\includegraphics[clip,width=\textwidth]{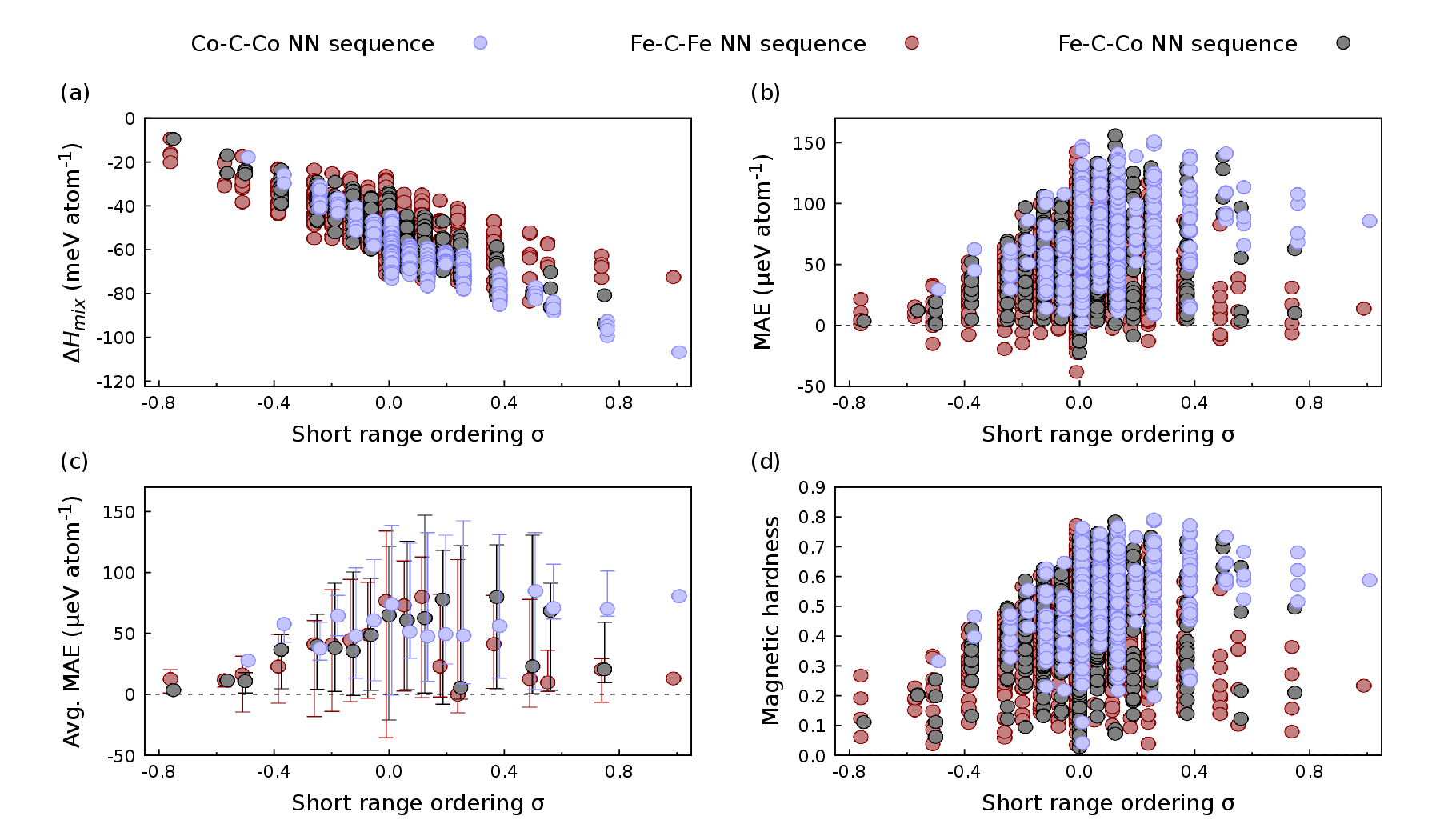}
	\caption{\label{fig-enthalpy-mae-hardness-vs-distance}
 Dependence of mixing enthalpy (a), magnetocrystalline anisotropy energy (b, c), and magnetic hardness (d) on short-range ordering parameter $\sigma$ in \fecoc{} structures with from 3 to 11 Co atoms in the supercell. The light blue color denotes systems with two Co atoms neighboring the C dopant, the dark red color indicates systems with two Fe atoms neighboring the impurity, and the black color denotes systems with the C atom neighbored by one Fe and one Co atom. Results were obtained using the FPLO18 code with PBE exchange-correlation potential. Fe-C-Fe and Co-C-Co data points are slightly shifted for better readability. Panel c presents thermodynamic averages according to Eq.~\ref{eqn-averaging}, and error bars denote the maximum and minimum calculated values.}
\end{figure*}

Considering different atomic configurations for particular Co concentrations makes it possible to determine the effect of the former on the values of magnetic moments on individual atoms.
D\'iaz-Ortiz~\etal{} shown for \feco{} that the average magnetic moment does not change significantly with ordering~\cite{diaz-ortiz_structure_2006}.
Similarly, for special quasirandom structures (SQS), comparing C impurity local neighborhood, Delczeg-Czirjak~\etal{} did not find any relevant change of total magnetic moment in the Fe\nobreakdash-Co\nobreakdash-C system.
The magnetic moment for the (Fe$_{0.5}$Co$_{0.5}$)$_4$C in their work remained at around 1.8 \mubat{}~\cite{delczeg-czirjak_stabilization_2014}.
Indeed, for \fecoc{}, we do not see any significant change in the average spin magnetic moment with the local chemical neighborhood, as shown in~Fig.~\ref{fig-mom-vs-ordering-histogram-enthalpy}(g--i).
Only a slight increase in the average spin magnetic moment with short-range ordering can be observed for the Fe$_8$Co$_8$C system, presented in~Fig.~\ref{fig-mom-vs-ordering-histogram-enthalpy}(h).
It validates effective medium approaches, such as VCA and CPA, to work for disordered \fecoc{}, similarly to \feco{}, the latter pointed by D\'iaz-Ortiz~\etal{}~\cite{diaz-ortiz_structure_2006}.
As for the average Fe and Co magnetic moments, we can see the variation across different structures drops with short-range ordering, indicating a strong contribution from Fe\nobreakdash-Co NN interaction.
It is consistent with a known strong Fe\nobreakdash-Co $d$ orbital hybridization and exchange interaction~\cite{odkhuu_firstprinciples_2019}. 
For any specific minority atom concentration in our computational cell, the $\sigma$ range is restricted due to limitations induced by the composition and system size, as described above.

\subsection{Ordering and its influence on magnetic properties}

Apart from average magnetic moment dependence on the short-range ordering, we can explore the ordering effect on other important system properties, including mixing enthalpy, magnetocrystalline anisotropy energy, and magnetic hardness.
Figure~\ref{fig-enthalpy-mae-hardness-vs-distance} presents aggregated results for Co content between 3 and 11 atoms in the system, in the bct region.
We do not present results for lower Co concentrations because they cover only a small number of configurations and do not have reasonable statistics.


Figure~\ref{fig-enthalpy-mae-hardness-vs-distance}(a) shows mixing enthalpy decrease with the increase of short-range Fe--Co ordering, i.e., the fraction of Fe--Co pairs among all NN pairs.
It might indicate system stabilization by Fe--Co nearest neighbor and Co--Co or Fe--Fe next-nearest neighbor interaction.
As previous studies have shown, in the case of the N\nobreakdash-doped B2 phase, nearest neighbors Fe--Co exchange integral and next-nearest neighbors Co--Co integral calculated by Odkhuu and Hong contribute the most to magnetic ordering~\cite{odkhuu_firstprinciples_2019}.
Hence, we ascribe the system stabilization to the same interactions.

 
Figure~\ref{fig-enthalpy-mae-hardness-vs-distance}(b) shows the distribution of MAE in structures with different atomic configurations.
Both the highest and lowest MAE for a single configuration can be observed for $\sigma$ equal to 0.
For the highest $\sigma$ values, MAE converges to around 85~\muevat{} for the Co\nobreakdash-C\nobreakdash-Co NN sequence and to around 10~\muevat{} for the Fe\nobreakdash-C\nobreakdash-Fe NN sequence.
For negative $\sigma$ values, which can be associated with low Co concentrations, the uniaxial MAE vanishes.
It can be deduced that the NN ordering influences the MAE by strong Fe\nobreakdash-Co interplay.
Nevertheless, the factor that contributes most to the overall behavior of the MAE relative to order is the direct immediate chemical neighborhood of the impurity atom.
In~Fig.~\ref{fig-enthalpy-mae-hardness-vs-distance}(c), we present thermodynamic averages, according to Eq.~\ref{eqn-averaging}.
Bars represent the range of MAE values obtained in calculations.
We observe no significant correlation between average MAE and atom distribution for $\sigma$ > 0.
The most probable MAE for $\sigma$ equal to 0 is quite high regardless of the dopant neighborhood.
The changes in MAE described above are clear, though the scatter of MAE values for various individual structures is substantial.

Taking all the above into account, the configuration space of Fe\nobreakdash-Co\nobreakdash-C alloys can be somewhat effectively reduced to random nearest-neighbor patterns. 
Still, it should be done cautiously and can lead to substantial errors, though any anomalies should be evident in the results.
Along with the low average magnetic moment dependence discussed above, the lack of strong MAE dependence on the short-range ordering implies that Fe\nobreakdash-Co\nobreakdash-C retains the properties of a random alloy, similarly to pure Fe\nobreakdash-Co.
Thus, methods relying on conformational space reduction by neighbor patterns analysis, such as SQS, yield a non-negligible error, similar to effective medium methods, as noted before by D\'iaz-Ortiz~\etal{}~\cite{diaz-ortiz_structure_2006}.
In future studies, it should be decided on a case-by-case basis whether the trade-off between the significant reduction in computation time in approximate (SQS-type) methods and the accuracy and ability to obtain a complete picture of the system in methods that allow order-dependence analysis is justified.

Figure~\ref{fig-enthalpy-mae-hardness-vs-distance}(d) presents a similar picture for magnetic hardness.
We can see that practical magnetic hardness can be obtained for systems around and above $\sigma$ = 0.
For highly ordered systems, the first coordination shell of the dopant plays a key part.
Above $\sigma$ = 0.4, only Co\nobreakdash-C\nobreakdash-Co and part of Fe\nobreakdash-C\nobreakdash-Co systems retain magnetic hardness in the semi-hard region.
The interesting part is the negative-$\sigma$ side of~Fig.~\ref{fig-enthalpy-mae-hardness-vs-distance}(b--d).
We observe that where Fe--Fe and Co--Co interactions dominate, MAE and hence the magnetic hardness drops.


\begin{figure}[ht!]
\centering
	\includegraphics[clip,width=\columnwidth]{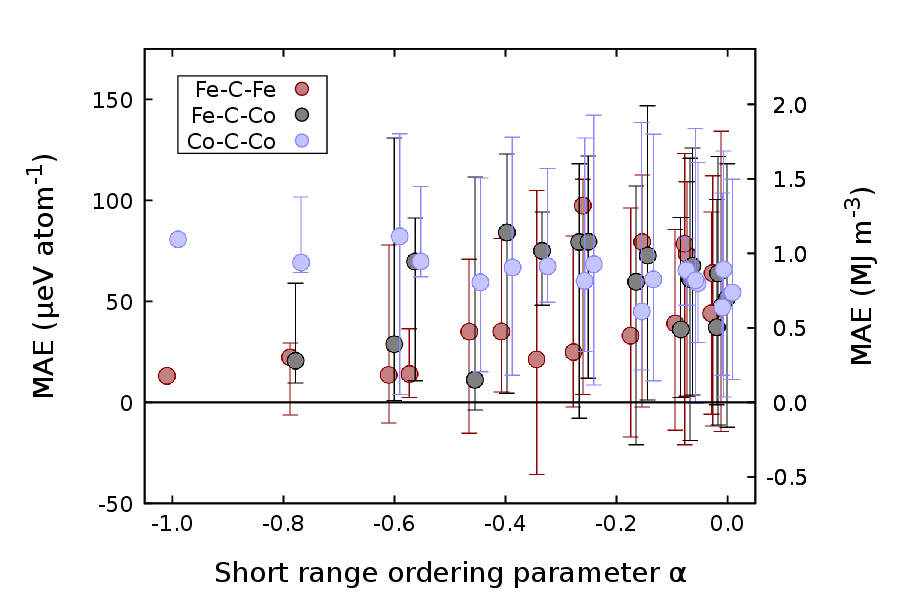}
	\caption{\label{fig-mae-vs-alpha}
        Magnetocrystalline anisotropy energy \textit{versus} Warren-Cowley short-range order parameter for the first coordination shell of transition metal atoms in \fecoc{} (from 3 to 11 Co atoms in the supercell). Points represent 300~K thermodynamical averages according to Eq.~\ref{eqn-averaging}. Results were obtained using the FPLO18 code with PBE exchange-correlation potential. Fe-C-Fe and Co-C-Co data points are slightly shifted for better readability. Bars denote the minimum and maximum values obtained in calculations.}
\end{figure}

Although the $\sigma$ is a convenient and effective parameter in analyzing the aggregated results, especially showing the linear decrease of mixing enthalpy with increasing dislike atom pairs content in the supercells, it lacks one property necessary to conduct a complete analysis.
It conveys a strict order parameter definition only for equiatomic binary alloy.
Namely, its expected value, the same as the value for a completely disordered alloy, is not always equal to zero and depends on minority atoms concentration $c_m$ as $4(c_m - c_m^2)$.
For equiatomic alloy, $\sigma$ equals 0 for completely disordered alloy and takes values up to 1 (or -1) for completely ordered alloys.

To investigate the properties of disordered alloys in a broad concentration range, we use Warren-Cowley short-range order parameter $\alpha$~\cite{cowley_approximate_1950,cowley_shortrange_1965}, which for the first coordination shell ($\alpha^{Fe,Co}_{I}$ -- shortened further to $\alpha$) can be simplified as:

\begin{equation}
\label{eqn-warren-cowley}
\alpha^{\rm AB}_{I} = 1 - \frac{p_{\rm AB}}{2 c_{\rm A} c_{\rm B}},
\end{equation}

\noindent where $c_{\rm A}$ denotes the concentration of atom type A, $p_{\rm AB}/2c_{\rm B} = P_{\rm AB}$ equals the conditional probability of finding an atom of type B at the first coordination shell of the randomly selected atom of type A, and when substituted, gives the exact Warren-Cowley formulation.
Structures with all $\alpha$ parameters (for different coordination shells) equal to 0 are disordered, and structures with $\alpha_i$ equal to 1 (or -1) are perfectly ordered on coordination shell $i$.
For simplicity, in Fig.~\ref{fig-mae-vs-alpha}, we present only MAE \textit{versus} $\alpha$ dependency.
Generally, in an infinite crystal, $\alpha$ takes values between $\frac{2 c_{\rm A} c_{\rm B} - 1}{2 c_{\rm A} c_{\rm B}}$ and 1~\cite{owen_new_2016}.
We get only zero to negative $\alpha$ values due to the small computational cell size.
Overall, the plot is similar to the positive $\sigma$ part of Fig.~\ref{fig-enthalpy-mae-hardness-vs-distance}(b) and (c) taking into account that preferred dislike atom type coordination is associated with positive $\sigma$, but negative $\alpha$.
The most probable MAE value is proportional to the ordering for Co\nobreakdash-C\nobreakdash-Co systems and, to some extent, for others.
Apart from that, we want to highlight three main observations.
Firstly, there is a considerable spread in values for random alloys (for $\alpha = 0$).
It is further indicator implying that certain methods of configurational space reduction, like SQS, are inherently predestined to fail in proper \feco{}-based alloys MAE predictions, and the uncertainty of such results can be, in fact, substantial.
Secondly, same as for $\sigma$ and similarly to order parameters in recent works by Izardar and Ederer for L1$_0$ FeNi~\cite{izardar_impact_2022,izardar_interplay_2020}, the MAE value converges towards a reasonably high MAE value for perfectly ordered systems.
Lastly, in all (Fe\nobreakdash-C\nobreakdash-Fe, Fe\nobreakdash-C\nobreakdash-Co, and Co\nobreakdash-C\nobreakdash-Co) systems, there is a group of configurations that possess high MAE, increasing with ordering.
We remind here that $\alpha = -1$ structures are ordered.
For Fe\nobreakdash-C\nobreakdash-Fe and Fe\nobreakdash-C\nobreakdash-Co systems, the average MAE value diverges and eventually suddenly drops for high-order structures -- a behavior described above for Bethe $\sigma$ dependencies.


From the comparison of high-order structures calculations to the random \feco{} alloy, D\'iaz-Ortiz~\etal{} deduced that ordered structures are stable, with B2 phase among them~\cite{diaz-ortiz_structure_2006}. 
Structures predicted by them, namely D0$_3$, L6$_0$, and B2, as well as similar phases such as L1$_2$ exhibit a high degree of short-range $\sigma$ and $\alpha$ ordering, as calculated according to Eqs.~\ref{eqn-bethe}~and~\ref{eqn-warren-cowley}.
Wu~\etal{} have, similarly, reported stability of Fe-rich D0$_3$, and equiatomic B2 phases~\cite{wu_firstprinciples_2008}, and Odkhuu and Hong postulated B2 Fe\nobreakdash-Co to be a good matrix for low-energy high-MAE N\nobreakdash-doped phases~\cite{odkhuu_firstprinciples_2019}.
One of the very first works on the topic of strained \feco{} system treated with CPA effective medium approximation by Turek~\etal{} researched \loz{} ordering influence on MAE in the system~\cite{turek_magnetic_2012}.
\loz{} and B2 phases differ only by lattice parameters $c/a$ ratio, where \loz{} is an fcc-like structure and B2 is close to bcc.
As such, we also checked specifically the B2 ordering in the low $c/a$ regime for the C\nobreakdash-doped \feco{} alloy.

For this purpose, we use the long-range order parameter $S$ of a binary alloy, which is defined in relation to a specific structure, in our case -- B2-like Fe$_8$Co$_8$C.
Ordering towards B2 and its equivalent \loz{} phase has been studied in VCA and CPA approaches in several works to date, including one by Turek~\etal{}~\cite{turek_magnetic_2012}.
The parameter $S$ value equal to 1 is associated with a perfect ordering towards the chosen structure (in our case -- an ideal crystal in the B2 type), and $S$ equal to 0 represents an absolute lack of the ordering of the given type.
Importantly, a system without ordering towards one structure can be perfectly ordered towards another structure, such as L1$_2$ structure having a zero $S$ towards L1$_0$, both being highly-ordered fcc-like structures and having a high degree of nearest-neighbor ordering.
Long-range ordering parameter $S$ can be represented in general as follows~\cite{bethe_statistical_1935,bragg_effect_1934,bragg_effect_1935,williams_effect_1935}:

\begin{equation}
	S = \frac{p - p(S=0)}{p(S=1) - p(S=0)},
\end{equation}

\noindent where $p$ denotes the probability of finding an atom of a given type on the expected atomic site. 
For two-atom type 2 $\times$ 2 $\times$ 2 supercell and B2 ordering we expand it as:

\begin{figure}[t]
\centering
	\includegraphics[clip,width=0.6\columnwidth]{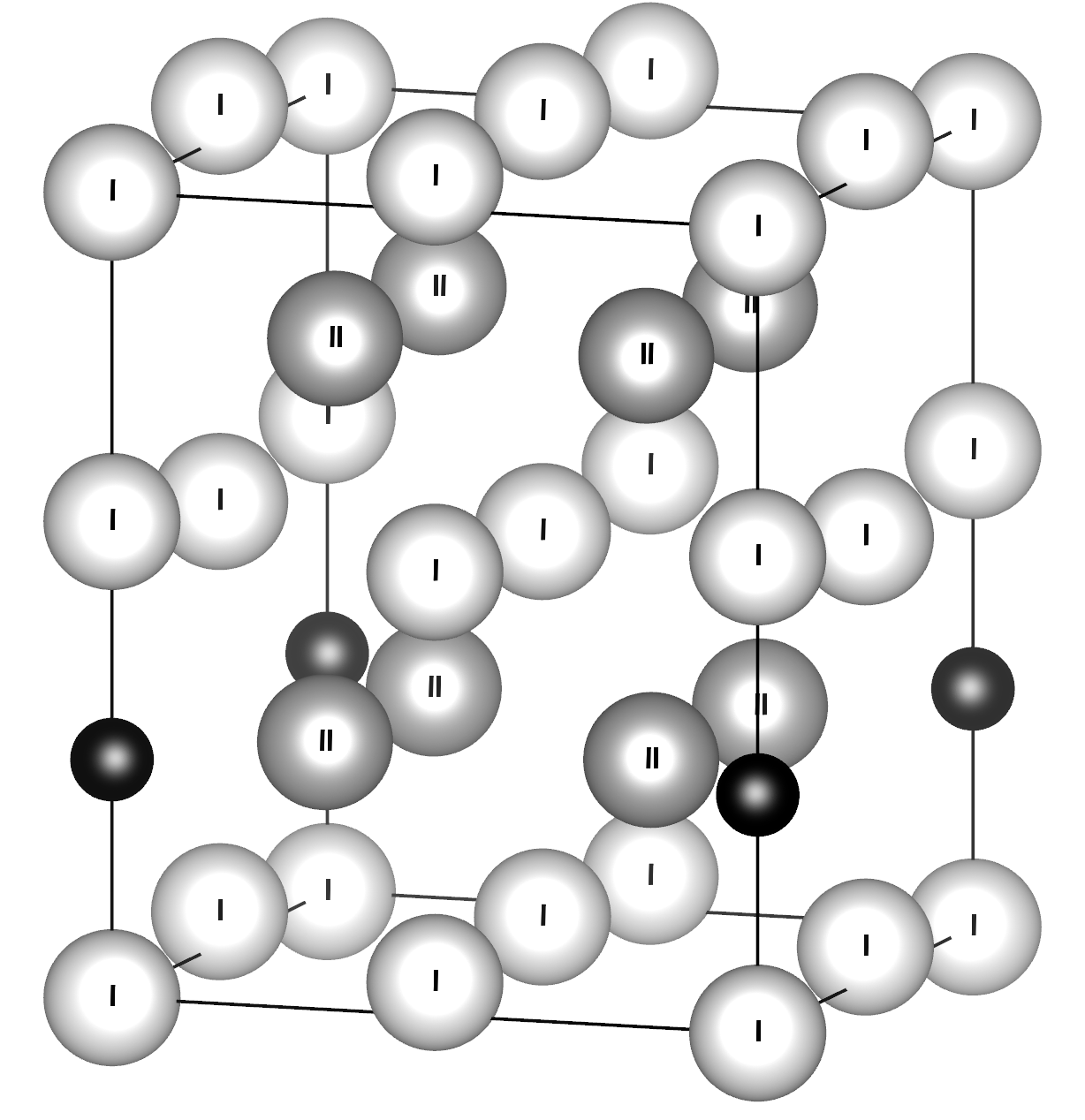}
	\caption{\label{fig-I-and-II}
Graphical representation of atomic sites relevant in Eq.~\ref{eqn-S-practical}. The presented structure is a 2~$\times$~2~$\times$~2 supercell with a single octahedral C atom (black). Sites I are located close to $z = 0$ or $z = 0.5c$ plane, and sites II lie close to $z = 0.25c$ or $z = 0.75c$ plane.}
\end{figure}

\begin{equation}
\label{eqn-S-practical}
	S = \frac{|N_{\rm I} - N_{\rm II}|}{N},
\end{equation}

\begin{figure*}[t!]
\centering
	\includegraphics[clip,width=\textwidth]{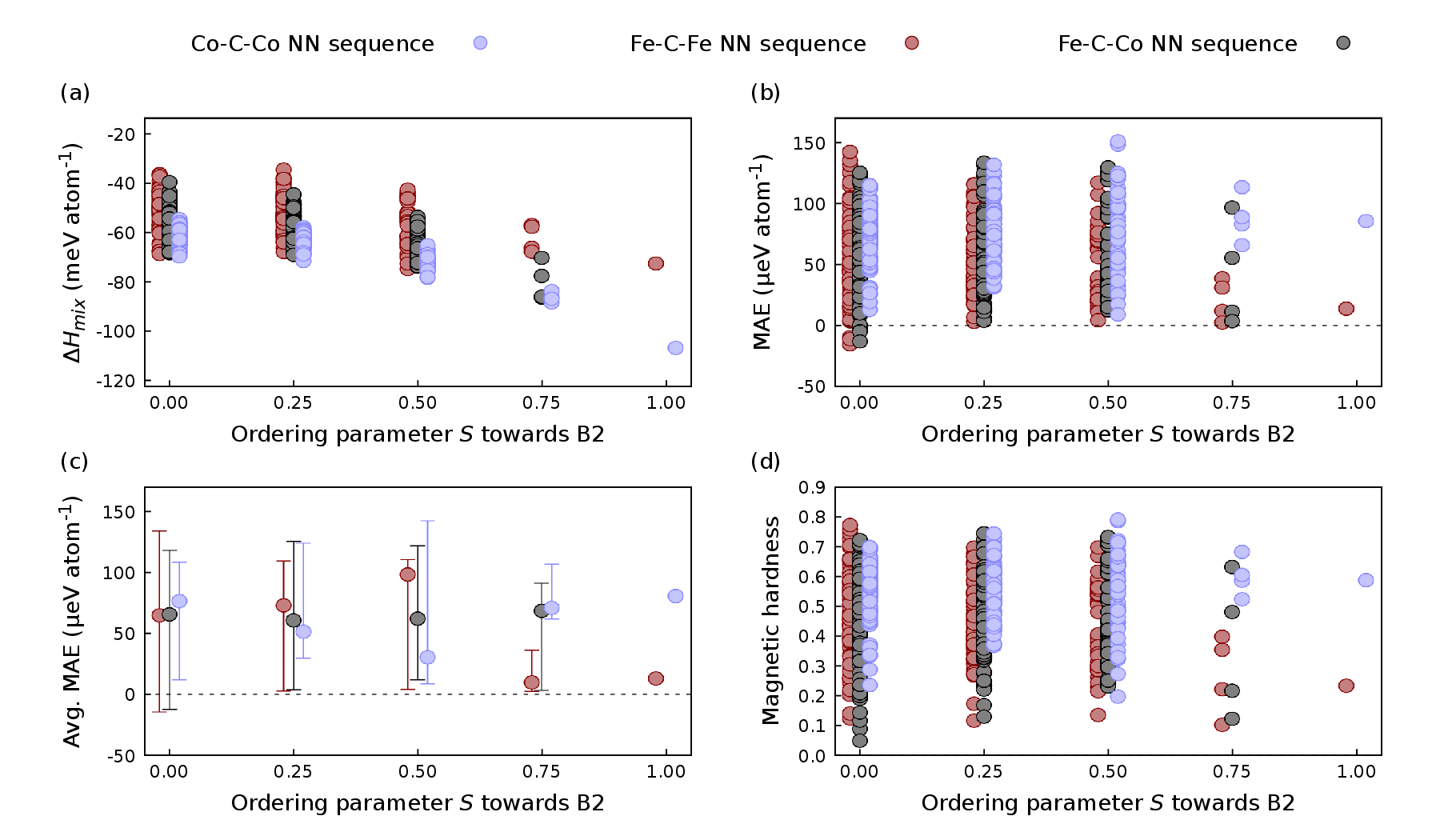}
	\caption{\label{fig-enthalpy-mae-hardness-vs-ordering}
 Dependence of mixing enthalpy (a), magnetocrystalline anisotropy energy (b, c), and magnetic hardness (d) on long-range ordering parameter $S$ in Fe$_8$Co$_8$C. The light blue color denotes systems with two Co atoms neighboring the C dopant, the dark red color indicates systems with two Fe atoms neighboring the impurity, and the black color denotes systems with the C atom neighbored by one Fe and one Co atom. Results were obtained using the FPLO18 code with PBE exchange-correlation potential. Fe-C-Fe and Co-C-Co data points are slightly shifted for better readability. Error bars on the c panel denote maximum and minimum calculated values.}
\end{figure*}

\noindent where $N_{\rm I}$ denotes the number of minority atoms close to $z = 0$ or $z = 0.5c$ plane, $N_{\rm II}$ denotes the number of minority atoms close to $z = 0.25c$ or $z = 0.75c$ plane, and $N$ is the sum of minority atoms in the system.
The sites are visualized in Fig.~\ref{fig-I-and-II}.
An effectively similar approach has been used recently by Izardar~\etal{} studying equiatomic FeNi L$\rm 1_0$ binary phase~\cite{izardar_interplay_2020,izardar_impact_2022}.
Parameter $S$ provides a linear scale, similar to one applied by Turek~\etal{}~\cite{turek_magnetic_2012}.


In Fig.~\ref{fig-enthalpy-mae-hardness-vs-ordering}, we show ordering towards B2 structure dependencies analogous to Fig.~\ref{fig-enthalpy-mae-hardness-vs-distance}, presenting results for short-range ordering parameter $\sigma$.
As the $S$ parameter towards B2 considers only equiatomic systems, the results aggregated are for Fe$_8$Co$_8$C only.
Similarly to $\sigma$-dependency, Fig.~\ref{fig-enthalpy-mae-hardness-vs-ordering}(a) presents a monotonic decrease in mixing enthalpy with B2 ordering in Fe$_8$Co$_8$C.
The energy of configurations with the Co\nobreakdash-C\nobreakdash-Co NN sequence is, on average, significantly lower than the energy of configurations with the Fe\nobreakdash-C\nobreakdash-Co NN sequence, which is, in turn, lower than the energy of Fe\nobreakdash-C\nobreakdash-Fe systems.
This fact is independent of the ordering.
Perfectly ordered B2 structure with C dopant between two Co atoms possesses the lowest energy.

\begin{figure*}[ht]
\centering
	\includegraphics[clip,width=\textwidth]{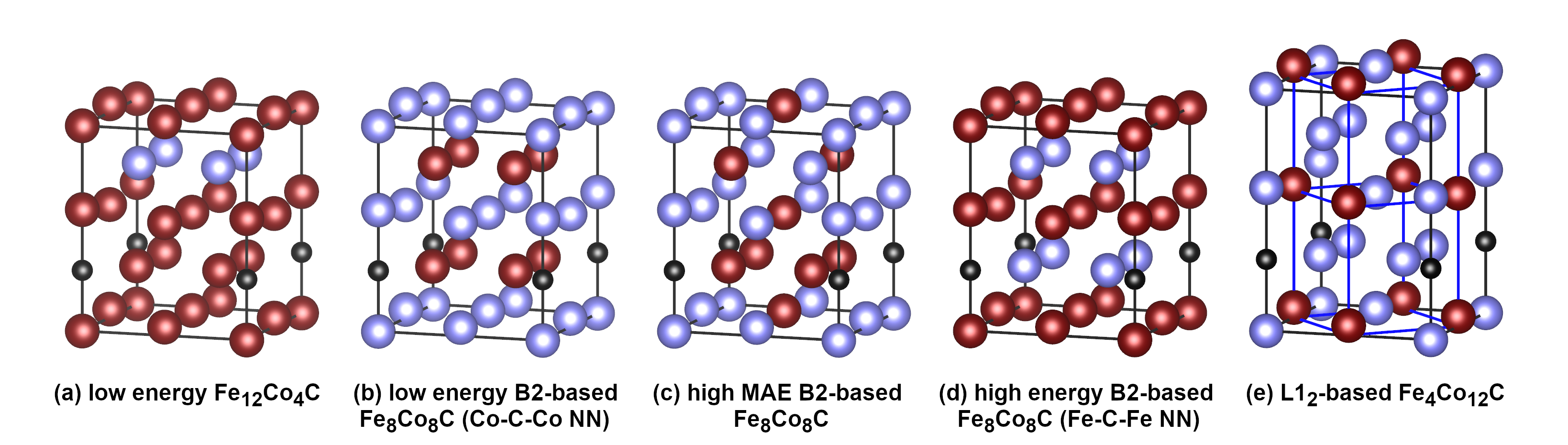}
	\caption{\label{fig-structures-2}
	Panels (a--c) present exemplary obtained low-energy, high-MAE, and high-symmetry supercells: Co interlayer separated from C impurity by half of the supercell (a), Co-C-Co B2 (b), and highest MAE Fe$_8$Co$_8$C (c). Panel (d) presents the high-energy Fe-C-Fe B2 structure, and panel (e) presents low-energy, high-MAE Fe$_4$Co$_{12}$C L$1_2$. In panel (e), the alternative fcc representation of the L$1_2$ structure is presented with blue lines. Supercell lattice parameters were optimized in FPLO/PBE with virtual crystal approximation, and atomic positions were optimized for a few steps in every atomic position occupancy.}
\end{figure*}

In Fig.~\ref{fig-enthalpy-mae-hardness-vs-ordering}(b), we see multiple atomic configurations deviating vastly from the average.
In fact, the single highest MAE value, which is twice the average, can be observed for $S$ = 0.5.
The associated structure is presented in Fig.~\ref{fig-structures-2}(c).
The qualitative agreement of MAE averages, presented in Fig.~\ref{fig-enthalpy-mae-hardness-vs-ordering}(c), with the work of Turek~\etal{} is good.
We can see that MAE does not follow any specific trend with B2 ordering.
For low ordering towards the B2 phase, we can see both very high and very low MAE values.
MAE value converges towards a reasonably high 85~\muevat{} for perfect B2 order and Co\nobreakdash-C\nobreakdash-Co configuration.
Conversely, for C impurity in the Co plane (neighbored by two Fe atoms), MAE converges towards a low value of approximately 10~\muevat{}.
These are exactly the same MAE values as for most positive sigma and most negative alpha parameters, see Figs.~\ref{fig-enthalpy-mae-hardness-vs-distance} and~\ref{fig-mae-vs-alpha}.
It is, in fact, the same structure, visualized further in Fig.~\ref{fig-structures-2}.
Magnetic hardness \textit{versus} B2 ordering, shown in~Fig.~\ref{fig-enthalpy-mae-hardness-vs-ordering}(d), have to be similar to MAE since the system magnetization has been shown above to not depend on the ordering.
The main conclusion is that for higher ordering, only systems with Co\nobreakdash-C\nobreakdash-Co and Fe\nobreakdash-C\nobreakdash-Co NN sequences possess practical magnetic hardness.
Similarly to $\sigma$, for low B2 ordering, we can still observe many individual atomic arrangements with hardness above 0.5.


It might be tempting to dive more deeply into the evaluated atomic occupation configurations individually, with particular emphasis on the high-symmetry structures.
However, such analysis is beyond the scope of this work, as we rely on error cancellation due to the high sample count.
A detailed look at the specific structures would require a much finer \textbf{k}-point mesh and fine atomic positions optimization of such atomic arrangements.
Nevertheless, to emphasize possible further paths of Fe\nobreakdash-Co\nobreakdash-C system investigation, we present in Fig.~\ref{fig-structures-2} four selected low-energy, high-MAE structures (a, b, c, and e), as well as a high-energy, low-MAE, perfectly ordered B2 structure (d).
We found that high-order structures for as low as 25\% Co concentration can indicate practical magnetic properties.
Interestingly, the lowest energy structure for Fe$_{12}$Co$_4$C is the Co interlayer in the plane farthest away from the C impurity.
These structures can be promising candidates for future permanent magnets since the price of Fe is negligible in the overall price of an Fe\nobreakdash-Co alloy.
As for qualitative trends, we observe the L1$_2$ structure among the lowest energy systems for high Co concentrations in the fcc regime.
Despite the structure changes towards bct with lowering of the Co content, the atomic occupations for low-energy Fe$_{12}$Co$_4$C remain the same as in the high-Co L1$_2$ phase, presented in Fig.~\ref{fig-structures-2}(e).

\section{Summary and Conclusions}


We conducted a full configuration space analysis for 2~$\times$~2~$\times$~2 \fecoc{} supercell based on a 2-atom body-centered tetragonal unit cell, with a single C impurity at one of the octahedral interstitial positions in the supercell.
The calculations were performed using density functional theory (DFT) with the generalized gradient approximation (GGA) using the full-potential local-orbit scheme (FPLO18).

In our tetragonal \fecoc{} supercells, we observe a structural phase transition from a body-centered tetragonal (bct) to a face-centered cubic (fcc) structure at a Co concentration of about 70 at.\%.
The lattice parameter $c/a$ ratio in the bct region ranges from 1.07 to 1.12.
We calculated relevant magnetic properties for all non-equivalent Fe/Co atom arrangements in the computational cell.
Since DFT calculations are, by definition, performed for a temperature of 0 K (for the ground state), we used thermodynamic averaging with an assumed temperature of 300 K in determining the average magnetocrystalline anisotropy energy (MAE) values.
Although, as previous experiments have shown, the structure expected above the critical Co concentration ($x \simeq 0.7$) is hexagonal, the assumed tetragonal geometry of the supercell does not allow this and leads to an fcc structure.

One of the basic features of the supercell geometry we analyzed is the first coordination shell of the C dopant atom.
The C atom has two nearest neighboring sites, which can be occupied by two Fe atoms, two Co atoms, or one Fe and one Co atom.
We found that for low Co concentrations, structures with impurities adjacent to two Fe atoms become more stable.
The expected result of the stabilization of the (Fe$_{0.5}$Co$_{0.5}$)$_X$C alloys by the Co\nobreakdash-C\nobreakdash-Co nearest neighbor sequence for medium to high Co concentrations is also confirmed in our results.

Although we observe a rather large spread of magnetic moments for different configurations on both Fe and Co atoms, the total magnetic moment in the supercell remains more or less constant.
Average (spin) magnetic moments decrease with increasing Co content, without a clear maximum for intermediate concentrations.

Positive MAE values in the bct region indicate a uniaxial magnetocrystalline anisotropy and show a broad maximum around medium Co concentration ($x \simeq 0.5$).
The calculated course of MAE as a function of Co concentration is in very good quantitative agreement with experimental data, which is a noteworthy improvement over effective medium methods.
The magnetic hardness of many configurations exceeds the conventional limit of 0.5 for magnetically semi-hard materials but does not exceed 0.9, remaining below the limit for hard magnetic materials.
In addition, for relatively low Co concentrations, on the order of 25\%, we have identified a number of energetically stable structures with high MAE values and potential economic significance.

The calculated mixing enthalpy of considered Fe\nobreakdash-Co\nobreakdash-C alloys is the lowest at around 50\% Co concentration.
Moreover, the general trends indicate that higher values of MAE (and magnetic hardness) correlate with more negative values of mixing enthalpy.
It shows that better structural stability coincides with high MAE.
Magnetocrystalline anisotropy energy values we obtain are comparable to the computationally derived MAE values of other C-, B-, and N-doped \feco{} systems, and slightly lower than both, experimentally and computationally obtained values for tetragonally strained undoped \feco{}.
Energy-wise, the C-doping only slightly shifts mixing enthalpy towards less preferable values, and the energetically worst-case atomic arrangements decline mixing enthalpy maximally by a factor of two.
It indicates maintaining relatively good structure stability and magnetic properties regardless of the atomic configuration.

A significant part of the discussion is devoted to determining the effect of ordering on the magnetic properties of the compositions under consideration.
We focus on the Bethe and Warren-Cowley short-range ordering parameters and the ordering parameter towards the arbitrarily chosen B2 (CsCl) structure.
In the largest range of values of the Bethe short-range ordering parameter, its increase correlates with an increase in MAE, while for the highest values of the parameter (above 0.2), we no longer track correlation.
Furthermore, we observe no significant correlation between MAE and the value of the Warren-Cowley short-range ordering parameter and the ordering parameter towards the B2 structure.
The direct neighborhood of the impurity dominates MAE value dependencies.
On the contrary, we see a clear decrease in the value of the enthalpy of mixing (higher stability) as short-range and long-range ordering parameters increase.


In summary, we present a relatively simple and effective method for averaging multiple configurations to predict accurate MAE values for the Fe\nobreakdash-Co\nobreakdash-C system.
We show that the method can be made even more efficient by averaging a few percent of the most energetically favorable structures, with little loss in accuracy.
In addition, the Fe\nobreakdash-Co\nobreakdash-C system is a good matrix for further modifications (e.g., induction of additional stresses) stabilized by the Fe\nobreakdash-Co nearest neighbor interactions.
Considering that B-, C-, and N\nobreakdash-doped Fe\nobreakdash-Co alloys possess similar structural and magnetic properties, further research of Fe/Co ordering in interstitially\nobreakdash-doped Fe\nobreakdash-Co can provide much-needed insight towards efficient, rare-earth-free permanent magnet development.
The utilized method allowed us, at a computational cost slightly exceeding that of the CPA method, to obtain basic magnetic characteristics with similar accuracy, and moreover, provided valuable information about the magnetic properties dependence on the internal structure of the alloy.

\section*{Acknowledgements}

We acknowledge the financial support of the National Science Centre Poland under the decision DEC-2018/30/E/ST3/00267. 
The work on the conformation-picking scheme development has been financed by the Polish Ministry of Science and Higher Education under the grant DI2017/007947.
Calculations were made in the Poznan Supercomputing and Networking Centre (PSNC/PCSS).
We acknowledge Paweł Leśniak for his help in compiling and maintaining the computational codes utilized.
We thank Joanna Marciniak and J\'an Rusz for their valuable discussion and suggestions.
We also want to thank Lucas Baldo, Karolina Olszewska, Justyna Rychły-Gruszecka, Justyn Snarski-Adamski, Maciej Szary, and Jan Raczyński for the valuable feedback on the manuscript.

\end{sloppypar}

\bibliography{FeCo-C}

\end{document}